\documentclass[11pt,english]{article}
\pdfoutput=1
\usepackage[T1]{fontenc}
\usepackage{fullpage}
\usepackage{authblk}
\usepackage{graphicx,array}
\usepackage{cite}
\usepackage{color}
\usepackage{caption}
\usepackage{latexsym}
\usepackage{amsthm}
\usepackage{amsmath}
\usepackage[titletoc]{appendix}
\usepackage{enumitem}
\usepackage{amssymb}
\usepackage{color}
\usepackage[unicode=true,
 bookmarks=true,bookmarksnumbered=false,bookmarksopen=false,
 breaklinks=false,pdfborder={0 0 1},backref=false,colorlinks=true]
 {hyperref}
\hypersetup{pdftitle={Cosmological Decoherence from Thermal Gravitons},
 pdfauthor={Ning Bao, Aidan Chatwin-Davies, Jason Pollack, Grant N. Remmen},
 citecolor=black,linkcolor=black,urlcolor=black}
\usepackage{breakurl}
\usepackage[hang,flushmargin]{footmisc} 
\usepackage{listings}
\usepackage{mathrsfs}
\usepackage{amsthm}
\usepackage{afterpage}

\newcommand{\Eq}[1]{Eq.~(\ref{#1})}
\newcommand{\Eqs}[2]{Eqs.~(\ref{#1}) and (\ref{#2})}
\newcommand{\Sec}[1]{Sec.~\ref{#1}}

\newcommand{\Fig}[1]{Fig.~\ref{#1}}
\newcommand{\App}[1]{App.~\ref{#1}}
\newcommand{\Ref}[1]{Ref.~\cite{#1}}

\usepackage{accents}

\newcommand{\bra}[1]{\langle #1 |}
\newcommand{\ket}[1]{| #1 \rangle}

\newtheorem{thm}{Theorem}[section]

\newtheorem{example}[thm]{Example}

\newcommand{\mbf}[1]{\mathbf{#1}}

\usepackage{accents}

\newcommand{\be}{\begin{equation}}
\newcommand{\ee}{\end{equation}}

\usepackage{xcolor}
\usepackage{color}
\definecolor{idealorange}{RGB}{251,156,49}
\definecolor{nicegreen}{rgb}{0.1,0.5,0.1}

\begin{document}

\interfootnotelinepenalty=10000
\hfill

\vspace{2cm}
\thispagestyle{empty}
\begin{center}
{\LARGE \bf
Cosmological Decoherence from Thermal Gravitons
}\\
\bigskip\vspace{1cm}{
{\large Ning Bao,${}^{a,b}$ Aidan Chatwin-Davies,${}^c$ Jason Pollack,${}^d$ and Grant N. Remmen${}^a$}
} \\[7mm]
 {\it ${}^a$Center for Theoretical Physics and Department of Physics \\
     University of California, Berkeley, CA 94720, USA and \\
     Lawrence Berkeley National Laboratory, Berkeley, CA 94720, USA \\[1.5 mm]
 ${}^b$Computational Science Initiative, Brookhaven National Lab, Upton, NY 11973, USA
 \\[1.5mm] ${}^c$KU Leuven, Institute for Theoretical Physics\\Celestijnenlaan 200D B-3001 Leuven, Belgium\\[1.5mm]
 ${}^d$Department of Physics and Astronomy\\University of British Columbia, Vancouver, BC V6T 1Z1, Canada} \let\thefootnote\relax\footnote{\noindent e-mail: \url{ningbao75@gmail.com}, \url{aidan.chatwindavies@kuleuven.be}, \url{jpollack@phas.ubc.ca}, \\ \hphantom{e-mail:} \url{grant.remmen@berkeley.edu}} \\
 \end{center}
\bigskip
\centerline{\large\bf Abstract}
\begin{quote} 
We study the effects of gravitationally-driven decoherence on tunneling processes associated with false vacuum decays, such as the Coleman--De~Luccia instanton. 
We compute the thermal graviton-induced decoherence rate for a wave function describing a perfect fluid of nonzero energy density in a finite region.
When the effective cosmological constant is positive, the thermal graviton background sourced by a de Sitter horizon provides an unavoidable decoherence effect, which may have important consequences for tunneling processes in cosmological history. 
We discuss generalizations and consequences of this effect and comment on its observability and applications to black hole physics.
\end{quote}
	
\setcounter{footnote}{0}

\newpage
\tableofcontents
\newpage
\baselineskip=18pt
\section{Introduction}
The phenomenon of decoherence \cite{Zeh:1970fop, Zurek:1981xq, Griffiths:1984rx,Joos:1984uk} is of great importance in characterizing the dynamics of our noisy universe. It has proven crucial to our understanding of quantum computing \cite{PhysRevA.52.R2493, PhysRevLett.75.3788}, black hole physics \cite{Bao:2017who, penington2019entanglement, almheiri2019entropy, almheiri2019page, almheiri2019islands,Agarwal:2019gjk,Piroli:2020dlx}, and quantum thermodynamics \cite{CH,2011RvMP...83..771C,TamirCohen}, among many other applications.

It is therefore a natural question to ask about the consequences of decoherence in general cosmological and gravitational settings. In particular, we will consider the effects of decoherence induced by a bath of thermal gravitons~\cite{Blencowe} on the physics of phase transitions, cosmological and otherwise. 
Since gravitons couple universally to all fields and cannot be screened without violating energy conditions, such decoherence is fundamentally unavoidable in gravitationally-interacting systems. Gravitionally-induced decoherence thus provides a floor for the decoherence rate in any physical system of potential relevance. While the gravitational decoherence rate may be quite small in the context of, for example, tabletop AMO experiments, these effects can, in fact, be physically significant in certain epochs of cosmological history. In this paper, we work in an effective field theory of general relativity~\cite{Donoghue:1994dn}, where the propagating graviton is obtained from perturbative expansion around a fixed background metric, and, following \Ref{Blencowe}, use the nonequilibrium quantum field-theoretic machinery of \Ref{CH} to assess the rate of gravitationally-driven decoherence for models of phase transitions such as Coleman--De~Luccia~\cite{Coleman:1980aw}. \Ref{Kiefer:2010pb} is the earlist work that we are aware of to study decoherence in Coleman--De~Luccia transitions, in which the decohering environment is modeled as a quadratically-coupled scalar field. More recently, \Ref{Bachlechner:2012dg} considered a decohering environment of thermal photons with a classical gravitational coupling to the Coleman--De~Luccia bubble. Some recent works that also consider generalizations of cosmological phase transitions beyond Coleman--De~Luccia are Refs.~\cite{Braden:2018tky,Blanco-Pillado:2019xny}.

The fact that gravitational decoherence is relevant to phase transitions in quantum field theory should not be a surprise. A key phenomenon in quantum field theory is the decay of some metastable state (particulate, vacuum, or otherwise). 
By definition, a system that decays is not in an eigenstate of the Hamiltonian, but instead has some decay width that scales inversely with its lifetime; that is, an unstable or metastable system is in a superposition of energy eigenstates.
Since the graviton couples to energy-momentum, if a system with a decay width is in contact with a thermal graviton bath, it must necessarily also experience gravitationally-sourced decoherence.

This paper is organized as follows. In \Sec{sec:PT}, we review how cosmological phase transitions can be treated as WKB processes, characterized by a field-space wave function, and define the regimes where gravitationally-driven decoherence can affect the tunneling rates and approximations used in such false-vacuum decay transitions. 
In \Sec{sec:decoherence} we set up our thermal calculation and write the master equation governing the evolution of the system's density matrix.
We then compute the noise integrals that set the decoherence rate for our system of interest in \Sec{sec:noise}.
We argue that the decoherence effect we derive is inherently quantum mechanical, in that it only appears when the thermal graviton ensemble correctly reflects the quantum statistical mechanics of a bosonic system. In particular, we show that the asymptotic decoherence rate disappears if the thermal graviton bath is treated classically. 
In \Sec{sec:evolution}, we analyze the time evolution of the system's density matrix, using the example of a two-level system to elucidate the salient points.
We find that the nature of the graviton-induced decoherence has a crucial dependence on the equation of state of the superposed $T_{\mu\nu}$; we argue in \Sec{sec:causality} that this dependence is unique to thermal graviton systems and is forbidden by unitarity and causality from occurring for thermal systems of lower spin.
We discuss further generalizations and consequences in \Sec{sec:discussion}, including possible connections to black hole physics, and conclude in \Sec{sec:conclusions}.

\section{Cosmological phase transitions}\label{sec:PT}

The setting that we will consider is a scalar field $\phi$ coupled to gravity and propagating in some potential $V(\phi)$.
The action takes the form
\be  \label{eq:action}
S = \int {\rm d}^4 x\sqrt{-g} \left[\frac{1}{2\kappa^2}R - \frac{1}{2}\partial_\mu \phi \partial^\mu \phi - V(\phi)\right],
\ee
where $\kappa^2 = 8 \pi G = m_{\rm Pl}^{-2}$, $R$ is the Ricci scalar, and we work in mostly-plus Lorentzian metric signature.
We are interested in particular in cases where $V(\phi)$ has multiple local minima, so that all but the lowest-potential minimum (or minima) are false vacua of the theory.\footnote{In fact, even the perturbative vacuum corresponding to the lowest minimum receives nonperturbative (instanton) corrections, which correct it to the actual lowest-energy eigenstate of the full potential, so over exponentially-long timescales we should expect perturbative excitations around this minimum to evolve to states with nonzero overlap with every perturbative minimum \cite{Boddy:2014eba}. On the timescales we consider, this effect should not be important, so we will follow usual practice in eliding this distinction and referring to the perturbative ground state of the lowest minimum as the true vacuum of the theory.}
While one can expand the field about any one of these vacua and study its excitations about this local minimum in the potential, these configurations will be unstable due to quantum tunneling to field configurations that are peaked about lower-potential vacua.
In other words, false vacua decay quantum mechanically towards the lowest-potential, true vacuum.

The physics of false vacuum decay was worked out by Coleman and Callan for scalar fields in Minkowski space \cite{Coleman:1977py,Callan:1977pt} and subsequently by Coleman and De Luccia for scalar fields coupled to gravity \cite{Coleman:1980aw}.
Consider a theory in which the potential has only one false vacuum $V_+$ at $\phi = \phi_+$ and one true vacuum $V_-$ at $\phi = \phi_-$.
The decay probability per unit time and per unit volume is given by
\begin{equation}
\Gamma/{\cal V} = A e^{-B/\hbar} \left[ 1 + O(\hbar) \right] ,
\end{equation} \label{eq:decayRateCdL}
where the coefficient $B$ is
\begin{equation}
B = S_E(\phi_*,g_*) - S_E(\phi_+,g_+),\label{eq:CdLB}
\end{equation}
writing $S_E$ for the analytic continuation of the action \eqref{eq:action} to Euclidean signature and $g_+$ for the metric that solves the (Euclidean) Einstein equations for a homogeneous field whose value is the false vacuum.
The pair $\phi_*$, $g_*$ are a nonconstant solution of the Euclidean Einstein equations that minimizes $S_E$, subject to the condition that $\phi_*$ approaches $\phi_+$ at Euclidean infinity.
In the case where the difference $V(\phi_+) - V(\phi_-)$ is small, Coleman and De Luccia were able to give closed-form expressions $B$, $\phi_*$, and $g_*$.
In particular, they argued that $\phi_*$ is spherically symmetric and that its profile consists of an approximately constant interior region with $\phi_* \approx \phi_-$, an approximately constant exterior region with $\phi_* \approx \phi_+$, and a ``thin wall'' that joins the two regions over which $\phi_*$ ramps up.
This instanton dominates false vacuum decay, resembling a bubble of true vacuum within the false vacuum.

The calculation of the decay rate $\Gamma/{\cal V}$ is essentially the product of a WKB approximation.
This is perhaps more clear in the quantum mechanical analogue of the field theory, a wave function of several continuous variables $\psi(\vec{q})$ under the influence of a potential $V(\vec{q})$.
According to a WKB approximation of the wave function in the vicinity of a minimum $\vec{q}_+$, the tunneling rate $\Gamma_\text{QM}$ is again given by an expression of the form
\begin{equation}
\Gamma_\text{QM} = a e^{-b/\hbar} \left[ 1 + O(\hbar) \right].
\end{equation}
Per the WKB approximation, the coefficient $b$ is given by \cite{Banks:1973ps,Banks:1974ij}
\begin{equation}
b = 2 \int_{\vec{q}_+}^{\vec{\sigma}} {\rm d}\vec{q} \sqrt{2 [V(\vec{q}) - V(\vec{q}_+)]}  ,
\end{equation}
where the integration is over a path from $\vec{q}_+$ to a point $\vec{\sigma}$ on the other side of the potential barrier around $\vec{q}_+$ for which $V(\vec{\sigma}) = V(\vec{q}_+)$ that minimizes the value of $b$.
It is then a relatively straightforward task to show that this expression for $b$ is equivalent to the value of the Euclidean action, $S_E$, evaluated on the extremal path $\vec{q}_*(\tau)$ with the boundary condition $\lim_{\tau \rightarrow \pm \infty} \vec{q}_*(\tau) = \vec{q}_+$ \cite{Coleman:1977py}.
For the bounce instanton in flat spacetime, an analogous calculation \cite{Coleman:1977py,Callan:1977pt,Coleman:1980aw} gives a value of $B$ that similarly depends on the shape of the potential, but that parametrically goes like 
\be 
B \sim \Delta \phi^4 (V_{\rm t} - V_0)^2/\Delta V^3,\label{eq:CdLestimate} 
\ee
where $V_{\rm t}$ is the characteristic size of the hilltop between $\phi_\pm$ and $\Delta \phi = |\phi_+ - \phi_-|$. Adding in the effect of gravity lowers $B$ by a multiplicative factor, somewhat accelerating decay, for the case of tunneling from a false to a true vacuum with positive cosmological constant~\cite{Coleman:1980aw}.

Depending on the shape of the potential, there is another instanton that can dominate, that of Hawking and Moss~\cite{Hawking:1981fz}.
In this instanton, the entire Hubble patch (in Euclidean space, the entire four-sphere) spontaneously tunnels to the hilltop between $\phi_+$ and $\phi_-$ where the potential is maximized, at some $\phi_{\rm t}$ (where we write $V_{\rm t} = V(\phi_{\rm t})$), and then subsequently rolls down to $\phi_\pm$.
In essence, one can view this instanton as a large, homogeneous thermal fluctuation of a Hubble patch up to the top of the potential barrier, followed by classical evolution to the minimum.
The Euclidean quantum gravity calculation again gives a tunneling rate as in \Eq{eq:decayRateCdL}, but where the exponent takes the form
\be 
B = \frac{24\pi^2}{\kappa^{4}}\left(\frac{1}{V_{\rm t}} - \frac{1}{V_+}\right),
\ee 
rather than the Coleman--De~Luccia form in \Eq{eq:CdLestimate}.

Our goal is to compare the decay rate of the false vacuum given by \Eq{eq:decayRateCdL} to the rate at which the wave functional of the scalar field decoheres when it is allowed to interact with a thermal bath of gravitons.
Thus we will need to write down and work with the actual wave functional of the field.
Doing so is quite difficult in general and remains so even in the context of the approximations used to study the Coleman--De Luccia instanton.
Therefore, we will further simplify the problem by restricting to a quantum theory of homogeneous field configurations.
An important distinction is that the original Coleman--De~Luccia calculation works in the approximation where both the scalar field and the metric degrees of freedom are taken to be ${\rm O}(4)$ symmetric in Euclidean signature; by computing the decoherence rate, we are relaxing one of these requirements, calculating the effect of the (necessarily inhomogeneous\footnote{Because the graviton distribution is necessarily inhomogenous, we might be concerned that some spatial configurations corresponding to fluctuations away from the thermal expectation might satisfy the Jeans criterion and hence lead to black hole formation. For sub-Planckian temperatures, however, the thermal energy in a small region is much smaller than that needed to fulfill the criterion; black hole formation from thermal fluctuations can only proceed if we have exponentially rare fluctuations that are either very energetic or on very large spatial scales. Hence we expect that we can neglect the rate of black hole formation compared to Coleman--De~Luccia tunneling. }) thermal graviton degrees of freedom, which do not show up in the Coleman--De~Luccia part of the path integral, on a background given by a homogeneous scalar field distribution.\footnote{One could treat the effects of thermal de~Sitter gravitons through higher-order terms in the path integral, so the effects we compute are formally subleading in $\hbar$ counting; we nonetheless find that the timescale relevant for decoherence will be short compared to the Coleman--De~Luccia timescale, indicating that the nonhomogeneous degrees of freedom are relevant for computing the effectively nonunitary evolution of the density matrix. Indeed, the derivation below will precisely compute the effects of these inhomogeneous degrees of freedom for the thermal bath. See Eqs.~\eqref{eq:PI} and \eqref{eq:SIF}. In general, inhomogeneous scalar degrees of freedom could also be present, and interactions in the scalar-field potential would also provide another source of decoherence. However, to remain maximally theory-independent, we restrict ourselves to computing only the gravitational source of decoherence, which is guaranteed to be present.}
The field's wave functional is quite tractable in this case, as its Schr\"odinger equation reduces to that of a single continuous degree of freedom.

Let us demonstrate this fact, for simplicity in a flat-spacetime context.
To regularize the problem, consider a quantized scalar field in a box of side length $L$ with periodic boundary conditions. (Later on, we will work instead in a spherical system of radius $L$, but for the present we will for clarity and simplicity consider the simple box; the modifications necessary for the spherical case are straightforward.)
Here we work in $3+1$ spacetime dimensions, but the same argument goes through for arbitrary dimension.
The forward and reverse Fourier transforms are
\begin{align}
\phi(\mbf{x}) &= \sum_{\mbf n} e^{i\omega_0 \mbf n \cdot \mbf x} \phi_{\mbf n}, \\
\phi_{\mbf{n}} &= \frac{1}{L^3} \int {\rm d}^3x~e^{-i\omega_0 \mbf n \cdot \mbf x} \phi(\mbf x) .
\end{align}
In the above, $\mbf{n} = (n_1, n_2, n_3) \in \mathbb{Z}^3$, $\omega_0 = 2\pi/L$, and the volume normalization has been chosen so that $\phi(\mbf x)$ and $\phi_{\mbf n}$ have the same mass dimension.

The Hamiltonian is given by
\begin{equation}
\label{eq:ham_orig}
H[\phi] = \int {\rm d}^3 x \left[ \frac{1}{2} \pi^2 + \frac{1}{2} \partial_i \phi \partial^i \phi + V(\phi) \right].
\end{equation}
Let us first momentarily ignore the potential $V(\phi)$, in which case it is straightforward to show that
\begin{equation}
\hat H = L^3 \sum_{\mbf n} \left[ \frac{1}{2} \hat \pi_{\mbf n} \hat \pi_{-\mbf n}     + \frac{1}{2} \omega_0^2 n^2 \hat \phi_{-\mbf n} \hat \phi_{\mbf n} \right].
\end{equation}
Next, from the canonical commutation relation
\begin{equation}
[\hat \phi(\mbf x), \hat \pi(\mbf y)] = i \delta^3(\mbf{x} - \mbf{y}) ,
\end{equation}
one finds that the commutation relation for individual modes $\hat \phi_{\mbf{n}}$ and $\hat \pi_{\mbf{n}'}$ is
\begin{equation}
[\hat \phi_{\mbf n}, \hat \pi_{\mbf{n}'}] = \frac{i \delta_{\mbf{n},-\mbf{n}'}}{L^3},
\end{equation}
for which a functional representation is given by
\begin{equation}
\hat \phi_{\mbf n} \equiv \phi_{\mbf n} \qquad \hat \pi_{\mbf n} \equiv \frac{-i}{L^3} \frac{\partial}{\partial \phi_{-\mbf n}} .
\end{equation}

Thus, the Hamiltonian comes to read
\begin{equation}
\hat H = \sum_{\mbf n} \left( -\frac{1}{2L^3} \frac{\partial}{\partial \phi_{-\mbf n}     } \frac{\partial}{\partial \phi_{\mbf n}}  + \frac{1}{2} L^3 \omega_0^2 n^2 \phi_{-\mbf n} \phi_{\mbf n} \right),
\end{equation}
and it acts on wave functions $\Psi(\{\phi_{\mbf n}\}_{\mbf n \in \mathbb{Z}^3})$ that possess a countable number of arguments.
If we now restrict to only the zero mode (i.e., we only consider constant field values), then the Hamiltonian simplifies to
\begin{equation}
\hat H = -\frac{1}{2L^3} \frac{\partial^2}{\partial \phi_0^2} + L^3 V(\phi_0) ,\label{eq:ham}
\end{equation}
where we have restored the potential $V$ in this special case.\footnote{When $V$ depends on arbitrary powers of $\phi$ and we keep all modes, then the expression for $V$ in terms of the Fourier modes is quite complicated.
However, keeping only the zero mode, the term $\int {\rm d}^3x~V(\phi)$ just boils down to $L^3 V(\phi_0)$.}

\begin{figure}[t]
\begin{center}
\includegraphics[width=0.5\columnwidth]{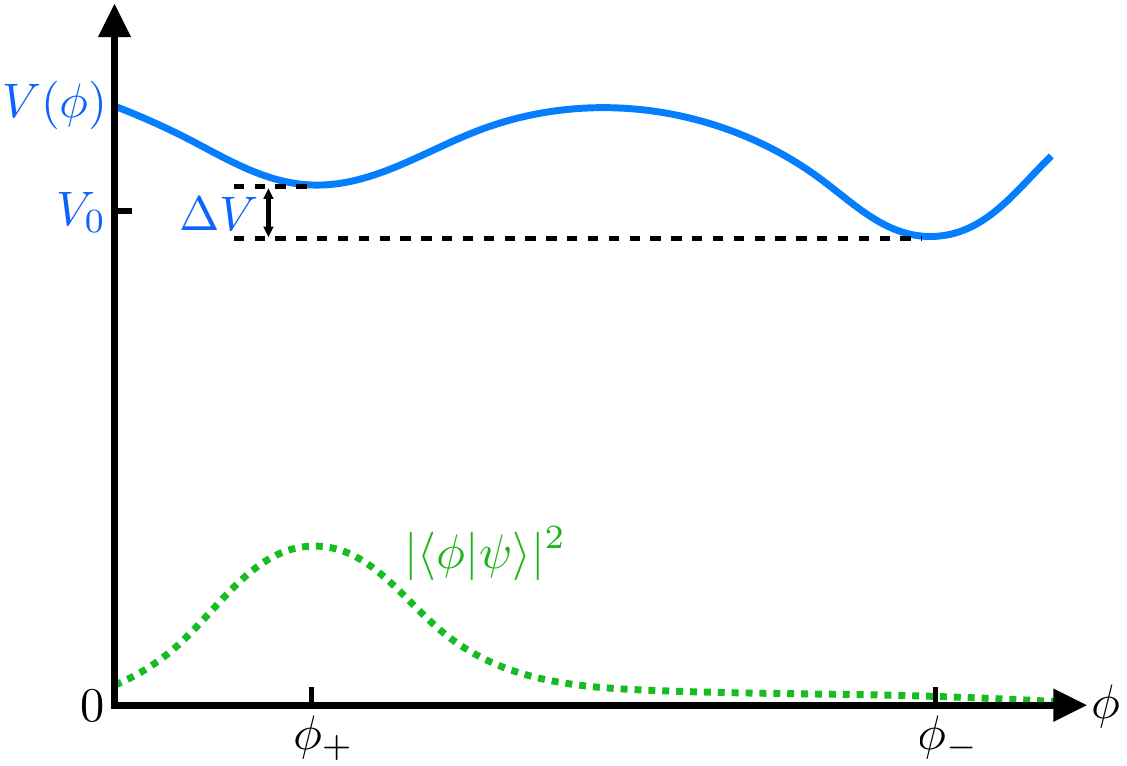}
\end{center}
\vspace{-6mm}
\caption{Potential with a false vacuum $\phi_+$ and true vacuum $\phi_-$, for which $\Delta V \ll V_0$ (solid blue curve). The state $|\psi\rangle$ of our system is not an eigenstate of $\phi$, but rather is characterized by some distribution with finite width in the $\phi$ basis. An example low-energy state initialized primarily in the false vacuum is depicted by the dashed green curve.}
\label{fig:potential}
\end{figure}

Hence, the field value for $\phi$ obeys the single-particle Schr\"odinger equation for a particle in a potential given by $V$; see \Fig{fig:potential}.
A reasonable starting point for our wave function in the false vacuum therefore approximates a Gaussian, with some exponentially-suppressed tail of support in the other vacuum, which sources the exponentially-suppressed decay process.
That is, our state is not an eigenstate of the Hamiltonian, but rather has a decay width.
Considering a low-energy wave function with support primarily near the false vacuum $\phi_+$, one can treat the Hamiltonian in \Eq{eq:ham} as a one-dimensional harmonic oscillator, expanding $V$ in $\varphi = \phi_0 - \phi_+$ as $V(\phi_+) + \frac{1}{2}V''(\phi_+)\varphi^2+\cdots$. Then the wave function in $\varphi$ goes as $\exp[-L^3 \sqrt{V''(\phi_+)} \varphi^2/2]$, which gives a characteristic variance in the total energy in our finite region, $\Delta E^2 \sim V''(\phi_+)$.

Let us define $\Delta V = [V(\phi_+)-V(\phi_-)]/2$ and $V_0 = [V(\phi_+) + V(\phi_-)]/2$. If $\Delta V \sim V_0$, then for general states with support near both vacua, it is not well-defined to ask what the background Hubble constant is.
However, if we consider a potential where $\Delta V$ is small, $\Delta V \ll V_0$, so that the ``thin wall'' approximation is valid, then in this limit both vacua have a background Hubble constant approximately dictated by $V_0$.
In particular, this background metric will source a de~Sitter temperature~\cite{Gibbons:1977mu}
\be
T_{\rm dS} = \frac{1}{2\pi} \sqrt{\frac{\kappa^2 V_0}{3}}.\label{eq:dStemp}
\ee
The de~Sitter horizon will source a thermal bath of all particle species in the theory, including thermal gravitons.
These gravitational degrees of freedom will provide an environment that couples to the superposition of $T_{\mu\nu}$ dictated by the $\phi$ wave function and will induce decoherence.\footnote{We note that away from the thin-wall approximation, the wall itself will have a complicated field profile that can act as another environment, and thus another competing source of decoherence, for the part of the wave function inside the false vacuum.}
It is this decoherence rate that we wish to compute, in order to see how it competes with the tunneling rate.\footnote{Since graviton-sourced decoherence is a perturbative process, we should generically expect it to be faster than the nonperturbative instanton effects of Coleman--De~Luccia and Hawking--Moss, and this is indeed what we will find below.}
If we have a quantum field theory with interactions for $\phi$ (e.g., the cubic term in the $\phi$ expansion about its false vacuum), then de~Sitter radiation in quantum fields other than gravitons will provide an additional source of decoherence.
However, as this would be a theory-specific calculation and we wish to remain theory-independent, we will restrict ourselves to computing the graviton-induced decoherence rate.
Due to the universal coupling and no-screening properties of gravity, this will provide an indelible source of noise that will give us a minimum estimate of the effects of decoherence.

Although all our calculations will be in the small $\Delta V$ limit, we briefly discuss the qualitative effects of more general potentials. First, even in cases where the potential differs by a significant amount at $\phi_+$ and $\phi_-$, we expect that one could meaningfully assign a de~Sitter temperature to the configuration, provided that the state is peaked only about one of the field values. For illustration, suppose that the state is peaked about a de Sitter vacuum at $\phi_+$, but susceptible to vacuum decay via bubble nucleation to $\phi_-$. Before the bubble nucleation, the spacetime thus has an ambient temperature corresponding to $\phi_+$, and when the bubble nucleates, particles at this temperature will monitor the nascent bubble. Hence, we expect that, for an initial wave function peaked around a false de Sitter vacuum, gravitational decoherence driven by thermal gravitons should still proceed. In particular, this observation should apply for Minkowski or anti-de~Sitter true vacua, which cannot even sensibly be assigned a graviton temperature. Of course, this will no longer be the case when considering more general calculations than the initial tunneling process, e.g.., in situations in which uptunneling is important and rate-equation computations involving long-time equilibrium states.
Moreover, since a decay from $V(\phi_+) \geq 0$ to $V(\phi_-) \leq 0$ necessarily has $\Delta V \geq V_0$, the deductive results that we obtain in the $\Delta V \ll V_0$ limit cannot be immediately ported over to more general vacuum decays.

Since we ultimately want a decoherence rate per unit volume, in analogy with \Eq{eq:decayRateCdL}, we will infrared-regulate the problem.
Motivated by the Coleman--De~Luccia instanton, we consider a region ${\cal B}$ of characteristic size $L$, where $L$ is of order the bubble radius, which for $\Delta V \ll V_0$ is much smaller than the Hubble radius.
Thus, for the purposes of computing the decoherence rate on ${\cal B}$, we can ignore the effects of the background spacetime curvature, and treat the problem as a system with a superposition of $T_{\mu\nu}$ in nearly-flat spacetime coupled to a thermal bath at temperature $T_{\rm dS}$.
To remain maximally theory-agnostic, we will not restrict ourselves to decays simply among different backgrounds of $w=-1$ (i.e., a superposition of different cosmological constants in ${\cal B}$), but will allow $T_{\mu\nu}$ to be described by any perfect fluid, thus allowing for more general consideration of the effect of decoherence on transitions between different field condensates \cite{Kampfer:2000gx,Kase:2018iwp,ArkaniHamed:2003uy,kanamoto2003quantum,fischer2009deconfinement,nagy2010dicke,Sin:1992bg}. For simplicity, we will consider a toy model of this more general transition, where the equation-of-state parameter is fixed, but where the wave function describes a distribution in overall density in ${\cal B}$.

\section{Gravitational decoherence}\label{sec:decoherence}

We now explicitly compute the effects of decoherence on our system.
In this section, we largely follow the formalism of Refs.~\cite{Blencowe,CH} but apply it to our cosmologically-relevant system of interest. Let us take as our starting point a massive scalar $\phi$ coupled to gravity as in \Eq{eq:action}, but where now for convenience we subtract off the constant term in $V(\phi)$ and write it as a cosmological constant by sending $R$ to $R-2\Lambda$.
We consider perturbations of the metric $g_{\mu\nu}$ around a geometry $\overline g_{\mu\nu}$ of constant spatial curvature, writing $g_{\mu \nu} = \overline g_{\mu \nu} + 2\kappa h_{\mu \nu}$ for a canonically normalized graviton. Expanding to quadratic order in the perturbation, we have
\be
S[h_{\mu \nu},\phi]=S_S[\phi] + S_E[h_{\mu \nu}] + S_I[h_{\mu \nu},\phi].
\ee
We are calling the term depending only on $\phi$ the action of the system,
\be 
S_S = \int{\rm d}^4 x\sqrt{-\overline g} \left[ -\frac{1}{2} \partial_\mu \phi \partial^\mu \phi - V(\phi)\right],\label{eq:SS}
\ee
and the term depending only on $h_{\mu\nu}$ the action of the environment,
\be 
\begin{aligned}
S_E &= \int{\rm d}^4 x\sqrt{-\overline g} \left[- \frac{1}{2} \nabla^\rho h^{\mu\nu}\nabla_\rho h_{\mu\nu} + \nabla_\nu h^{\mu\nu} \nabla^\rho h_{\mu\rho} - \nabla_\mu h \nabla_\nu h^{\mu\nu} + \frac{1}{2} \nabla^\mu h \nabla_\mu h \right.
\\& \left.\qquad\qquad\qquad\qquad + \Lambda \left(h_{\mu\nu} h^{\mu\nu} - \frac{1}{2} h^2 \right) + \frac{\Lambda}{\kappa^2}\right],
\end{aligned}\label{eq:SE}
\ee
in anticipation of integrating out the environment to determine the effective evolution of $\phi$ alone.
In Eqs.~\eqref{eq:SS} and \eqref{eq:SE}, raising of indices and covariant derivatives are defined with respect to the background metric $\overline g_{\mu\nu}$ and $h = \overline g^{\mu\nu} h_{\mu\nu}$.
Adding the appropriate Faddeev--Popov gauge-fixing term for harmonic gauge in a curved background~\cite{Cheung:2016say}, $S_{\rm gf} = -\int{\rm d}^4 x\sqrt{-\overline g} F_\mu F^\mu$ for $F_\mu = \nabla^\nu h_{\mu\nu} - \frac{1}{2} \nabla_\mu h$, we have, up to a total derivative,
\be
S_E + S_{\rm gf} = \int{\rm d}^4 x\sqrt{-\overline g} \left[ \frac{1}{2}h_{\mu\nu}(\Box + 2\Lambda)h^{\mu\nu} - \frac{1}{4} h(\Box + 2\Lambda) h) + \frac{\Lambda}{\kappa^2} \right], \label{eq:SEgf}
\ee
where $\Box = \nabla_\mu \nabla^\mu$.
The interaction term is 
\be
S_I = \int {\rm d}^4 x\sqrt{-\overline g} \left(\kappa T^{\mu\nu} h_{\mu\nu} + \kappa^2 U^{\mu\nu\rho\sigma} h_{\mu\nu}h_{\rho\sigma}\right),\label{eq:SI}
\ee
where 
\be
T_{\mu\nu} = \partial_\mu \phi \partial_\nu \phi - \frac{1}{2}\overline g_{\mu\nu} \partial_\rho \phi \partial^\rho \phi - \overline g_{\mu\nu} V(\phi)
\ee
and, in analogy with the flat-space result of \Ref{Arteaga:2003we},
\be
U_{\mu\nu\alpha\beta} = -2 \overline g_{\nu\alpha}\partial_\mu \phi \partial_\beta \phi + \overline g_{\mu\nu} \partial_\alpha \phi \partial_\beta \phi + \left(\frac{1}{2} \overline g_{\mu\alpha}\overline g_{\nu\beta} - \frac{1}{4} \overline g_{\mu\nu} \overline g_{\alpha\beta}\right)[\partial^\sigma \phi \partial_\sigma\phi + 2 V(\phi)].
\ee

Now, the reduced state of the system is given by evolving the initial state and then tracing out the graviton environment~\cite{Blencowe,CH,Feldman:1991nr}:
\be
\rho_S[\phi,\phi^\prime,t]=\int {\rm d}h_{\mu\nu} \rho[\phi h_{\mu\nu}, \phi^\prime h^\prime_{\mu\nu}, t] = \int {\rm d}h_{\mu\nu} e^{i S[\phi,h_{\mu\nu}]} \rho[\phi h_{\mu\nu}, \phi^\prime h^\prime_{\mu\nu}, 0] e^{-i S[\phi^\prime, h^\prime_{\mu\nu}]},\label{eq:reduced_exact}
\ee
where $\rho_S[\phi,\phi^\prime,t] = \bra{\phi}\hat{\rho}(t)\ket{\phi^\prime}$, etc.
If the initial state is uncorrelated between the two fields,\footnote{The assumption of an initial product state, where system and environment are initially uncorrelated, is a standard one in the decoherence literature and is very natural in an idealized experimental context where a system is first prepared in isolation and then exposed to an environment. It may seem less natural in a cosmological setting. We will ultimately be interested only in the master equation describing the change of the density matrix under infinitesimal time evolution, so the assumption on the initial state is largely a matter of convenience. Physically, we merely need to be in a regime where it makes sense to think of the system as distinct from the environment, far from the maximum-entropy state where the system has completely equilibrated with the environment, so that the entropy production required for decoherence is allowed \cite{Halliwell:1999xh,2012NJPh...14h3010J}. In our case, we know that we are in an excited state of a false vacuum with finite decay width, very far from equilibrium.
If the initial state is entangled but in this tractable regime, we may expand in a basis of product states and do the integration separately for each term, resulting in a sum over integrals with a different (state-dependent) influence action in each term and thus a sum over different noise and dissipation kernels in the master equation. One could do a perturbative analysis or, e.g., carry out an analysis with an initial state with a finite number of terms in superposition, but this is beyond the scope of our analysis.}
\be
\rho[\phi h_{\mu\nu}, \phi^\prime h^\prime_{\mu\nu}, 0] = \rho_S[\phi,\phi^\prime,0]\rho_E[h_{\mu\nu},h^\prime_{\mu\nu},0],\label{eq:PI}
\ee
we can explicitly do the integration over $h_{\mu\nu}$ in  \Eq{eq:reduced_exact}:
\be
\begin{aligned}
&\rho_S[\phi,\phi^\prime,t] \\
&= \int \rho_S[\phi,\phi^\prime,0] e^{i (S_S[\phi]-S_S[\phi^\prime])} \int_{h_{\mu\nu}} \rho_E[h_{\mu\nu},h^\prime_{\mu\nu},0] e^{i (S_E[h_{\mu\nu}] + S_I[h_{\mu\nu},\phi] - S_E[h_{\mu\nu}^\prime] - S_I[h^\prime_{\mu\nu},\phi_\prime])} \\
&\equiv \int \rho_S[\phi,\phi^\prime,0] e^{i (S_S[\phi]-S_S[\phi^\prime]+S_{IF}[\phi,\phi^\prime])},\label{eq:SIF}
\end{aligned}
\ee
where we have defined the Feynman-Vernon influence action $S_{IF}$ (for a review, see, e.g., Sec.~3.2 of Ref.~\cite{CH}). 
Differentiation with respect to $t$ and Taylor expansion of $S_{IF}$ to lowest nontrivial order then yields the master equation for the system.
Since we will be interested in systems with characteristic length scale much smaller than the Hubble scale set by $\Lambda^{-1/2}$, we can apply the approximately-Minkowski results of \Ref{Blencowe}, writing the master equation to leading order in $\kappa$ as\footnote{The coupling of $h^2$ to the $U$ tensor in \Eq{eq:SI} would only contribute to the four-point function in $\phi$ at loop order (i.e., higher order in $m_{\rm Pl}$) and hence is negligible for our calculation, as in Ref.~\cite{Blencowe}.}
\be
\begin{aligned}
\partial_t \rho(t) &= \! -i [H, \rho(t)] - \!\int_0^t \!{\rm d} t'\! \int\!{\rm d}{\bf r}\!\int\!{\rm d}{\bf r}' \bigg[ N({\bf r} - {\bf r}',t') \Big(2 [T_{\mu\nu}({\bf r}),[T^{\mu\nu}({\bf r}',-t'),\rho(t)]]\\
&\qquad\qquad\qquad\qquad\qquad\qquad\qquad\qquad\qquad\qquad\; -[T_\mu^{\;\;\mu}({\bf r}),[T_\nu^{\;\;\nu}({\bf r}',-t'),\rho(t)]]\Big)\\
&\qquad\qquad\qquad\qquad\qquad\qquad\qquad\;\; -i D({\bf r} - {\bf r}',t') \Big(2 [T_{\mu\nu}({\bf r}),\{T^{\mu\nu}({\bf r}',-t'),\rho(t)\}]\\
&\qquad\qquad\qquad\qquad\qquad\qquad\qquad\qquad\qquad\qquad\; -[T_\mu^{\;\;\mu}({\bf r}),\{T_\nu^{\;\;\nu}({\bf r}',-t'),\rho(t)\}]\Big)\! \bigg],\label{eq:master}
\end{aligned}
\ee
where $H$ is the Hamiltonian of the system itself, i.e., of the field $\phi$, and $N$ and $D$ are the noise and dissipation kernels,
\be
\begin{aligned}
N({\bf r},t) &= \frac{\kappa^2}{8} \int\frac{{\rm d}{\bf k}}{(2\pi)^3} \frac{e^{i{\bf k}\cdot {\bf r}}}{k} \cos(k t)\left[n(k)+\frac{1}{2}\right]\\
D({\bf r},t) &=\frac{\kappa^2}{16} \int\frac{{\rm d}{\bf k}}{(2\pi)^3} \frac{e^{i{\bf k}\cdot {\bf r}}}{k} \sin(k t),
\end{aligned} \label{eq:NDdef}
\ee
where $n(k)$ is the occupation number per degree of freedom at temperature $T=\beta^{-1}$. Since gravitons are spin-two, quantum mechanical particles obeying Bose--Einstein statistics, we have
\be
n(k) = (e^{\beta k} - 1)^{-1}. \label{eq:BoseEinstein}
\ee
As shown in \Ref{CH}, the noise kernel involves an integral over the energies of the modes, weighted by thermal occupation number $n(k)$, along with a zero-point energy represented by the $+1/2$ in \Eq{eq:NDdef}.

As discussed in \Sec{sec:PT}, we will be interested in states where $T_{\mu\nu}$ is independent of position.
Our state for $T_{\mu\nu}$ is described by a perfect fluid with fixed equation-of-state parameter $w$ and with density $\Phi$ (corresponding to some field value $\phi$), so that in a local Lorentz frame, we have
\be
T_{\mu\nu} = \Phi \times {\rm diag}(1,w,\ldots,w). \label{eq:Tmunu}
\ee
Our Hilbert space is therefore characterized by the different values of $\Phi$. The system is coupled to an environment of thermal gravitons.
We will introduce an infrared regulator for the integrals in our noise and dissipation kernels: instead of separately integrating over ${\bf r}$ and ${\bf r}'$, we will change variables and integrate both ${\bf r} - {\bf r}'$ and ${\bf r}$ over a ball ${\cal B}$  of fixed radius $L$. (Any other infrared-regulation scheme for the Fourier transforms would just induce ${\cal O}(1)$ factors in the decoherence rate estimate and will not change our conclusions.)
Consequently, the Hamiltonian is $H = {\cal V} \Phi$, where we have written ${\cal V}$ for the volume of the ball ${\cal B}$.

The master equation \eqref{eq:master} then reduces to:
\be
\partial_t \rho(t) =  -i{\cal V}[\Phi,\rho(t)] - {\cal V} f(w) \Big({\cal N}(t)[\Phi,[\Phi,\rho(t)]]  - i {\cal D}(t)[\Phi,\{\Phi,\rho(t)\}] \Big),\label{eq:master2}
\ee
where 
\be
\begin{aligned}
{\cal N}(t) &= \int_0^t {\rm d}t' \int_{\cal B} {\rm d}{\bf r}\,N({\bf r},t')\\
{\cal D}(t) &= \int_0^t {\rm d}t' \int_{\cal B} {\rm d}{\bf r}\,D({\bf r},t'),
\end{aligned} \label{eq:ND}
\ee
and where we have defined the parameter
\be
f(w)= 1 - 3 w(w-2).
\ee
In order to concretely compute the evolution of the density matrix and find the rate of decoherence induced by thermal gravitons, we must now evaluate ${\cal N}$ and ${\cal D}$ in \Eq{eq:ND}.

\section{Thermal noise}\label{sec:noise}

In order to use the formalism of \Sec{sec:decoherence} to quantitatively characterize the evolution of the density matrix and in particular understand the late-time decoherence into the pointer basis, we must explicitly compute the functions ${\cal D}$ and ${\cal N}$ in \Eq{eq:ND}, which generate oscillatory and noise terms, respectively, in the master equation.
This calculation is performed in detail in \App{app:integrals}; we will state the results here. 

For the dissipation integral, we have
\be
{\cal D}(t) = 32\kappa^2 L^2 \left[1+(a^{2}-1)\theta(1-a)\right]\label{eq:calDfinal},
\ee
defining $a\equiv t/L$ and writing $\theta$ for the Heaviside step function. 

For ${\cal N}(t)$, it is physically informative to split the function up into the sum ${\cal N}_0(t)+{\cal N}_1(t)$, where ${\cal N}_0(t)$ is defined analogously with ${\cal N}$, but including only the zero-point (the $+1/2$) part of $N$ in \Eq{eq:NDdef}, while ${\cal N}_1$ takes the $n(k)$ part; see \App{app:integrals} for details. Defining $b\equiv\beta/L$, we find:
\be
\begin{aligned}
{\cal N}_0(t)&= \frac{\kappa^2 L^2}{8\pi} \left\{\frac{a}{2} + \frac{1}{8}(1-a)(1+a) \log\left[\left(\frac{1+a}{1-a} \right)^2 \right] \right\}\\
{\cal N}_1 (t) &= \frac{\kappa^2 L^2}{8\pi}\biggl\{ \frac{\pi}{3b}-\frac{a}{2}-\frac{1}{8}(1-a)(1+a)\log\left[\left(\frac{1+a}{1-a}\right)^{2}\right]\\
 & \qquad\qquad\qquad+\frac{b}{4\pi}\left[{\rm Li}_{2}\left(e^{-\frac{2\pi(1+a)}{b}}\right)+\mathfrak{R}{\rm Li}_{2}\left(e^{\frac{2\pi(1-a)}{b}}\right)\right]\\
 & \qquad\qquad\qquad+\frac{b^{2}}{8\pi^{2}}\left[{\rm Li}_{3}\left(e^{-\frac{2\pi(1+a)}{b}}\right)-\mathfrak{R}{\rm Li}_{3}\left(e^{\frac{2\pi(1-a)}{b}}\right)\right]\biggr\},
\end{aligned}\label{eq:calNN}
\ee
where we have used the polylogarithm function ${\rm Li}_n(z) = \sum_{n=1}^\infty z^k/k^n$ and have defined the real part of the dilogarithm and trilogarithm for all real $z>0$,
\be 
\begin{aligned}
\mathfrak{R}{\rm Li}_2(z) &= \begin{cases} {\rm Li}_2(z) &z\leq 1 \\ \frac{\pi^2}{3} - \frac{1}{2}(\log z)^2 - {\rm Li}_2(z^{-1}) & z>1  \end{cases}\\
\mathfrak{R}{\rm Li}_3(z) &= \begin{cases} {\rm Li}_3(z) & z\leq 1 \\ {\rm Li}_3(z^{-1}) + \frac{\pi^2}{3} \log z  - \frac{1}{6}(\log z)^3 & z>1 \end{cases},
\end{aligned}
\ee
which is real and continuous for all positive $z$. We thus have the nice expression
\be
\begin{aligned}
{\cal N}(t) &= \frac{\kappa^2 L^2}{8\pi}\biggl\{ \frac{\pi}{3b}+\frac{b}{4\pi}\left[{\rm Li}_{2}\left(e^{-\frac{2\pi(1+a)}{b}}\right)+\mathfrak{R}{\rm Li}_{2}\left(e^{\frac{2\pi(1-a)}{b}}\right)\right]\\
 & \qquad\qquad\;\;\;\;\,\,+\frac{b^{2}}{8\pi^{2}}\left[{\rm Li}_{3}\left(e^{-\frac{2\pi(1+a)}{b}}\right)-\mathfrak{R}{\rm Li}_{3}\left(e^{\frac{2\pi(1-a)}{b}}\right)\right]\biggr\}.
\end{aligned} \label{eq:calNfinal}
\ee

\begin{figure}[t]
\begin{center}
\includegraphics[width=0.7\columnwidth]{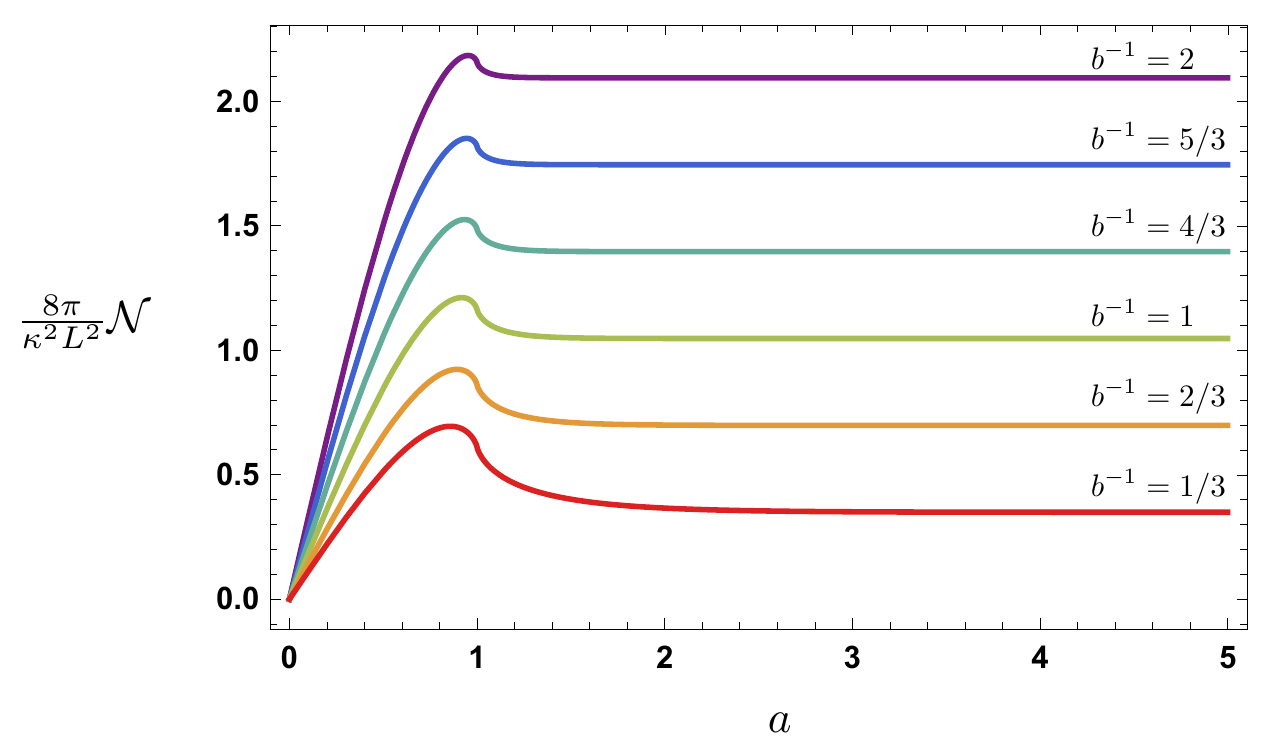}
\end{center}
\vspace{-6mm}
\caption{Noise function ${\cal N}$ from \Eq{eq:calNfinal}, written as a function of the rescaled time $a=t/L$, for several different values of the temperature parameterized by $b^{-1}=TL$. The asymptotic decoherence rate, set by the value of ${\cal N}$ at late times, is $\propto T$. }
\label{fig:N}
\end{figure}

\begin{figure}[t]
\begin{center}
\includegraphics[width=0.7\columnwidth]{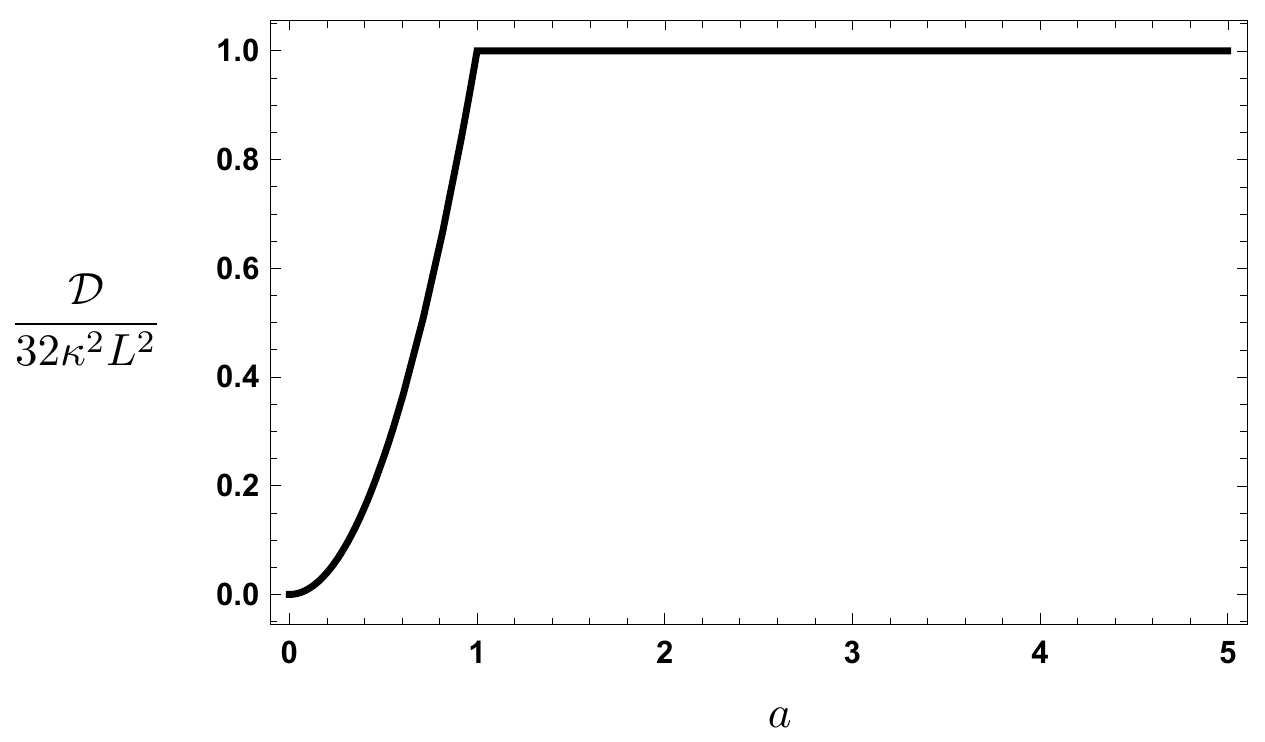}
\end{center}
\vspace{-6mm}
\caption{Dissipation function ${\cal D}$ from \Eq{eq:calDfinal} in terms of the rescaled time $a=t/L$.}
\label{fig:D}
\end{figure}

\begin{figure}[t]
\begin{center}
\includegraphics[width=0.75\columnwidth]{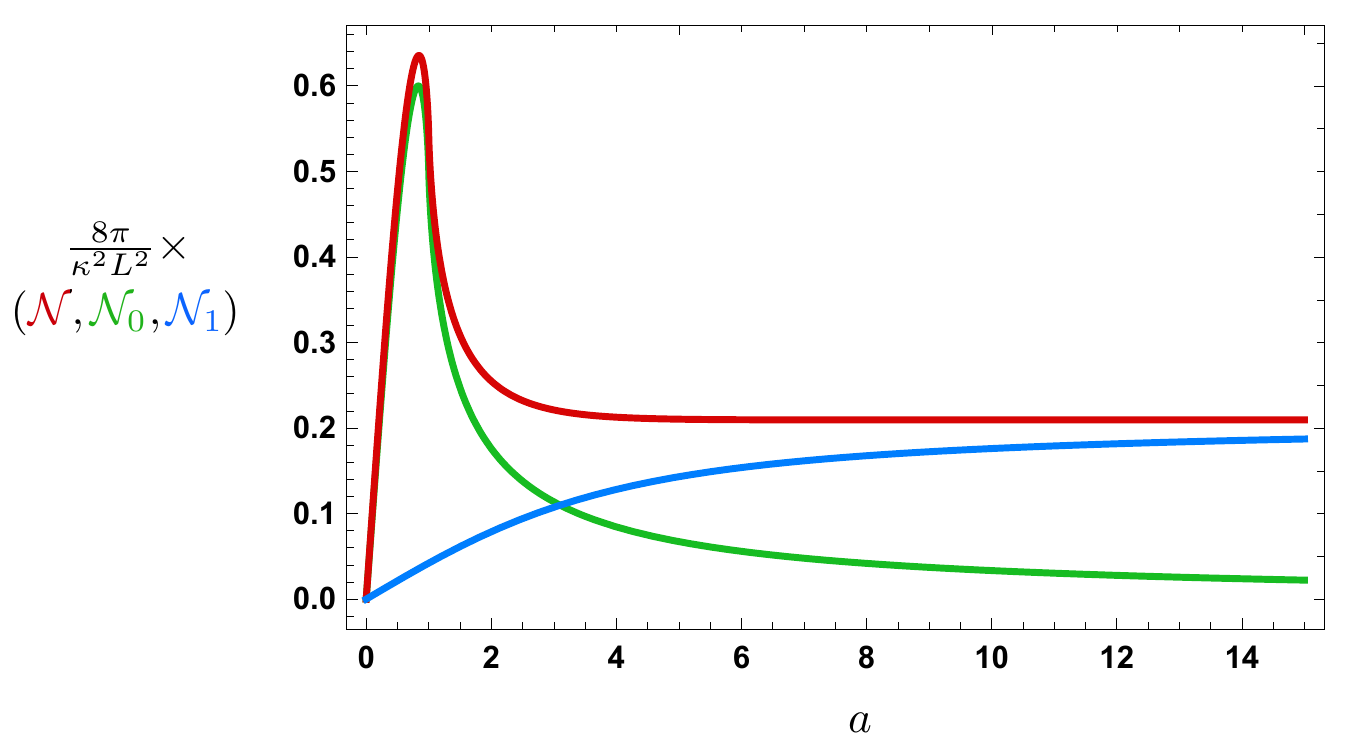}
\end{center}
\vspace{-6mm}
\caption{The noise function ${\cal N}$ from \Eq{eq:calNfinal} (red), written as the sum of its zero-point part ${\cal N}_0$ (green) and Bose--Einstein occupation number contribution ${\cal N}_1$ (blue) from \Eq{eq:calNN}, as a function of the rescaled time $a=t/L$. In this example, the functions are plotted for rescaled inverse temperature $b = 5$.}
\label{fig:NN}
\end{figure}

For timescales longer than the thermal timescale $\beta$ and the light-crossing time $L$, i.e., for $a \gg \max\{1,b\}$, only the $\pi/3b$ term in braces in \Eq{eq:calNfinal} survives, and we have
\be
{\cal N}(t) \xrightarrow {a \gg 1,b} \frac{\kappa^2 L^3}{24 \beta}.\label{eq:calNfinallimit}
\ee
See Figs.~\ref{fig:N} and \ref{fig:D} for an illustration of ${\cal N}$ and ${\cal D}$, respectively, and \Fig{fig:NN} for a breakdown of ${\cal N}$ into ${\cal N}_0$ and ${\cal N}_1$.

\subsection{Classical comparison}\label{sec:classical}

The noise term in \Eq{eq:calNfinal}, which sets the decoherence rate, was computed using the Bose--Einstein distribution \eqref{eq:BoseEinstein} for the occupation numbers of the thermal states.
It is instructive to compare this result with what we would have obtained for ${\cal N}(t)$ had we replaced $n(k) + \frac{1}{2}$ with the Maxwell--Boltzmann distribution,
\be 
n(k)_{\rm MB} = e^{-\beta k},\label{eq:MaxwellBoltzmann}
\ee
which specifies the thermal occupation numbers for a {\it classical} thermal ensemble.
That is, if our system is quantum mechanical, but it couples to a classical thermal noise source described by \Eq{eq:MaxwellBoltzmann}, does the system still decohere?

\begin{figure}[t]
\begin{center}
\includegraphics[width=0.75\columnwidth]{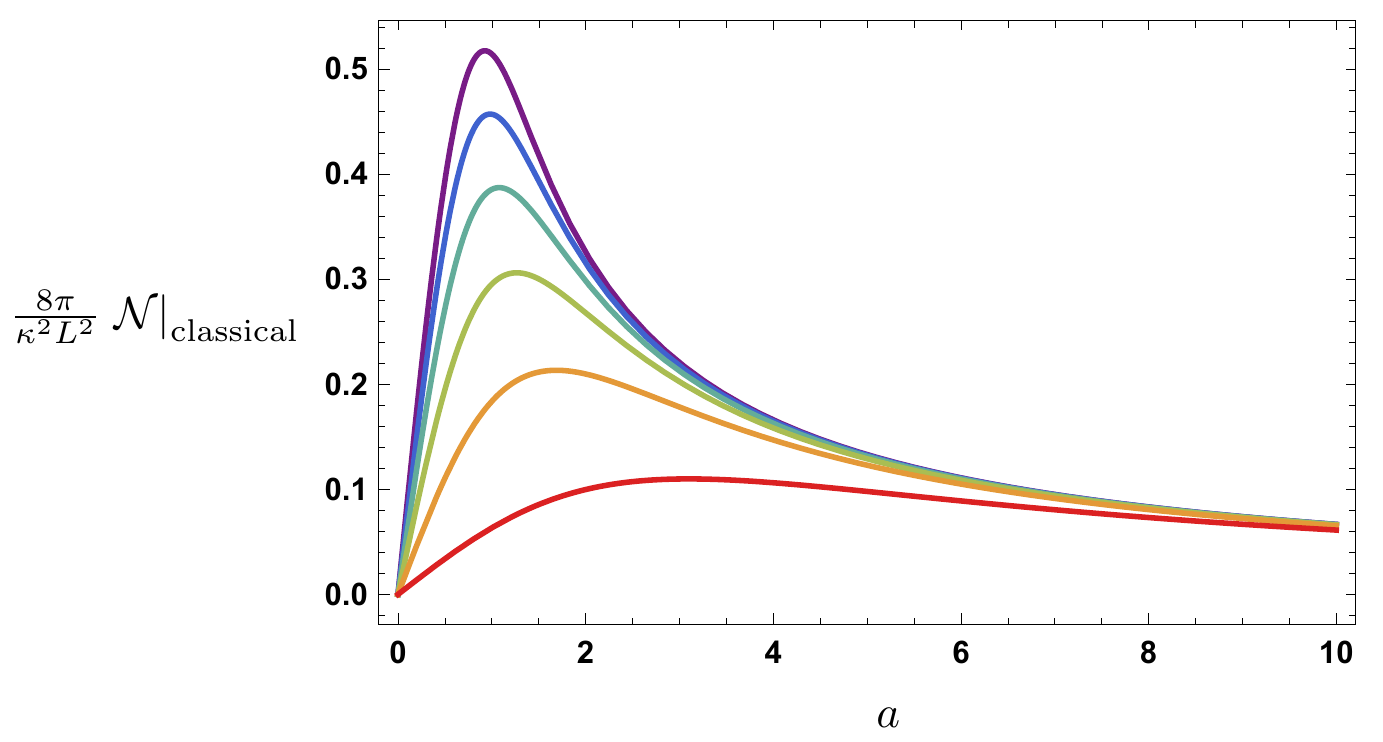}
\end{center}
\vspace{-6mm}
\caption{Noise function $\left.{\cal N}\right|_{\rm classical}$ from \Eq{eq:calNMB}, written as a function of the rescaled time $a=t/L$, for the same representative values of the temperature as in \Fig{fig:N}. The asymptotic decoherence rate, set by the value of ${\cal N}$ at late times, goes to zero as shown in \Eq{eq:noclassicaldecoherence}.}
\label{fig:Nclass}
\end{figure}

We can again compute ${\cal N}_1(t)$, with the only difference being the replacement in \Eq{eq:MaxwellBoltzmann}. The zero-point contribution from ${\cal N}_0(t)$ is dropped,\footnote{Note that the zero-point term ${\cal N}_0$ in \Eq{eq:calNN} does not contribute to the asymptotic decoherence rate anyway: ${\cal N}_0(t)\rightarrow 0$ for large $t$.} while  we find:
\be 
\begin{aligned}
L\int_0^\infty {\rm d}k\,\cos(kt)n(k)_{\rm MB} j_1(kL) &= 1 - \frac{a}{4} \log\left[\frac{(1+a)^2 + b^2}{(1-a)^2 + b^2} \right] \\&\qquad - \frac{b}{2} \left[\arctan\left(\frac{1+a}{b} \right) + \arctan\left(\frac{1-a}{b} \right) \right],
\end{aligned}
\ee
where $j_1$ is the first spherical Bessel function of the first kind. Performing the integration over $t$, we have: 
\be
\begin{aligned}
\left.{\cal N}(t)\right|_{\rm classical} &=\frac{\kappa^{2}L^2}{8\pi}\Biggl\{a+\frac{1}{4}(1+b^{2}-a^{2})\log\left[\frac{(1+a)^{2}+b^{2}}{(1-a)^{2}+b^{2}}\right]\\
 & \qquad\qquad\qquad-ab\left[\arctan\left(\frac{1+a}{b}\right)+\arctan\left(\frac{1-a}{b}\right)\right]\Biggr\}.
\end{aligned}\label{eq:calNMB}
\ee
For $t \gg \max\{L,\beta\}$, i.e., for timescales larger than the thermal transient and light-crossing time of the system, we have
\be
\left.{\cal N}(t)\right|_{\rm classical} \simeq \frac{\kappa^2 L^3}{12\pi t} \rightarrow 0,\label{eq:noclassicaldecoherence}
\ee
as shown in \Fig{fig:Nclass}. Thus, coupling a quantum system to a classical thermal noise source described by the Maxwell--Boltzmann distribution yields no long-term decoherence rate.\footnote{More concretely, a noise function ${\cal N}$ that goes like $1/t$ will lead to off-diagonal elements of the density matrix that decay as a power law, rather than exponentially, with time; see \Sec{sec:OD}.}
In other words, taking the $\beta \rightarrow 0$ limit of the thermal distribution does not commute with the decoherence integral over $k$.
The decoherence rate described by \Eq{eq:calNfinallimit} is an effect that relies on the Bose--Einstein distribution, i.e., the quantum nature of the environment: in this case, the fact that the thermal gravitons themselves are quantum mechanical.

\section{Evolution of the density matrix}\label{sec:evolution}

We will now apply the calculation of the noise and dissipation kernels from \Sec{sec:noise} to the master equation in \Eq{eq:master2} to compute the evolution of the density matrix, in order to characterize the pointer basis and decoherence of our stress-energy superposition.
We first consider an example two-level system.

\subsection{Off-diagonal decay}\label{sec:OD}

Let us consider a two-dimensional system Hilbert space, characterized by two states of $T_{\mu\nu}$ given by \Eq{eq:Tmunu} for fixed $w$, with $\Phi$ allowed to take two possible values, $\Phi_1$ and $\Phi_2$, corresponding to some field values $\phi_{1,2}$. Then $\rho$ is a two-by-two density matrix, which without loss of generality we can write as
\be 
\rho=\left(\begin{array}{cc}
\rho_{1} & \rho_{{\rm e}}\\
\rho_{{\rm e}}^{*} & 1-\rho_{1}
\end{array}\right),\label{eq:rho2level}
\ee
where $\rho_{1}$ is real.

We have:
\be 
\begin{aligned}
{} [H,\rho] &={\cal V}(\Phi_{1}-\Phi_{2})\left(\begin{array}{cc}
0 & +\rho_{{\rm e}}\\
-\rho_{{\rm e}}^{*} & 0
\end{array}\right)\\
[\Phi,[\Phi,\rho]]&=(\Phi_{1}-\Phi_{2})^{2}\left(\begin{array}{cc}
0 & +\rho_{{\rm e}}\\
+\rho_{{\rm e}}^{*} & 0
\end{array}\right)\\
[\Phi,\{\Phi,\rho\}]&=(\Phi_{1}+\Phi_{2})(\Phi_{1}-\Phi_{2})\left(\begin{array}{cc}
0 & +\rho_{{\rm e}}\\
-\rho_{{\rm e}}^{*} & 0
\end{array}\right),
\end{aligned}
\ee
so the master equation \eqref{eq:master2} reduces to the evolution of the off-diagonal element $\rho_{\rm e}$,
\be
\partial_{t}\log\rho_{{\rm e}}=-i{\cal V}(\Phi_{1}-\Phi_{2})-{\cal V}{\cal N}(t)(\Phi_{1}-\Phi_{2})^{2}f(w)+i{\cal V}{\cal D}(t)(\Phi_{1}+\Phi_{2})(\Phi_{1}-\Phi_{2})f(w),\label{eq:master3}
\ee
while $\rho_1$ is time invariant.

Since we are more interested in the asymptotic behavior of the density matrix, rather than thermal transients that occur on timescales of order $\beta$ or $L$, let us take replace ${\cal D}$ and ${\cal N}$ by their long-time behavior from \Eqs{eq:calDfinal}{eq:calNfinallimit}, so that
\be 
\partial_{t}\log\rho_{{\rm e}}\simeq -i{\cal V}(\Phi_{1}-\Phi_{2})-\frac{\kappa^2 L^3 {\cal V}}{24\beta}(\Phi_{1}-\Phi_{2})^{2}f(w)+32i\kappa^2 L^2  {\cal V} (\Phi_{1}+\Phi_{2})(\Phi_{1}-\Phi_{2})f(w).
\ee
Writing
\be
\rho_{\rm e}(t) = \rho_{\rm e,0} e^{-(\lambda+i\omega)t}
\ee
as an ansatz, we have
\be
\begin{aligned}
\lambda &= \frac{4\pi \kappa^2 L^6}{72\beta}f(w)(\Phi_1 - \Phi_2)^2 \\
\omega &= \frac{4}{3}\pi L^3 (\Phi_1 - \Phi_2) [1-32\kappa^2 L^2 f(w)(\Phi_1 + \Phi_2)],
\end{aligned}
\ee
where we substituted ${\cal V}=4\pi L^3/3$ for the volume of our system.

When $f(w)$ is positive, the states of definite $\phi$ (i.e., $\Phi=\Phi
_1$ or $\Phi_2$) form a pointer basis, and the off-diagonal term $\rho_{\rm e}$ characterizing their overlap decays away at a rate $\lambda$, which we can rewrite as
\be 
\lambda = \frac{T}{4}f(w)\left(\frac{E_1-E_2}{M_{\rm Pl}}\right)^2,\label{eq:lambda}
\ee
where $M_{\rm Pl} = 1/\sqrt{G}$ is the (unreduced) Planck mass and $E_{1,2} = {\cal V}\Phi_{1,2}$ are the two total energy states of the system.\footnote{If we had used the classical version of ${\cal N}$ from \Eq{eq:noclassicaldecoherence} instead, then $\rho_{\rm e}$ would decay as a power law, ${\mathfrak R}\,\rho_{\rm e} = \rho_{\rm e}(t=t_0) \times (t/t_0)^p$, where the power $p = f(w)(E_1 - E_2)^2/2\pi M_{\rm Pl}^2$, i.e., the decay is scale-free. In contrast, in the (correct) quantum case, the characteristic energy scale is simply $\lambda$.}
This decoherence occurs when $f(w)>0$, i.e., when 
\be
w \in \left(1-\frac{2}{\sqrt{3}},1+\frac{2}{\sqrt{3}} \right)\approx (-0.155, 2.155).
\ee
For $E_{1,2}$ fixed, the decoherence rate peaks for stiff matter, when $w=1$.

An equation of state $w>1$ is fairly exotic, breaking the dominant energy condition and thus leading to apparent acausal flow of energy-momentum in certain reference frames. On the other hand, taking $w<-0.155$ (though not less than $-1$) is still well motivated physically, with the salient example being the cosmological constant with $w=-1$. In such a situation, we have $f(w)<0$, so the off-diagonal element of the density matrix \eqref{eq:rho2level} {\it grows} with time, rather than shrinking. In this case we can remove the exponential time-dependence in the off-diagonal term by transforming to a new basis, writing 
\be 
\begin{aligned}
|\Psi_1\rangle &= \frac{1}{\sqrt{2}}\left(|\Phi_1\rangle - e^{i\omega t}|\Phi_2\rangle \right)\\
|\Psi_2 \rangle &= \frac{1}{\sqrt{2}}\left(e^{-i\omega t} |\Phi_1\rangle + |\Phi_2\rangle\right),
\end{aligned}
\ee
so that our basis states are those with superpositions of different $T_{\mu\nu}$. In the $|\Psi_{1,2}\rangle$ basis, the density matrix is
\be
\left. \rho \right|_{\Psi_{1,2} \text{ basis}}  =\left(\begin{array}{cc}
\frac{1}{2}-\rho_{{\rm e},0}e^{-\lambda t} & e^{-i\omega t}(\rho_1 - \frac{1}{2})\\
e^{i\omega t}(\rho_1 - \frac{1}{2}) & \frac{1}{2}+\rho_{{\rm e},0}e^{-\lambda t} 
\end{array}\right).
\ee
Recall that for $f(w)<0$, $\lambda$ is negative, so once $\lambda t \sim \log \rho_{{\rm e},0}$, the density matrix fails to be positive definite, indicating that the Born approximation in the master equation has broken down.
A particularly interesting choice is $\rho_1 = 1/2$, meaning that, in our definite-$\Phi$ basis, we have two equal entries on the diagonal (i.e., the two definite-$\Phi$ branches of the wave function have equal weight).
Then in the $\Psi$ basis, the off-diagonal elements then vanish, meaning that the $|\Psi_1\rangle$ and $|\Psi_2\rangle$ branches are entirely decohered.
As the wave function evolves, the $|\Psi_1\rangle$ branch has decaying weight, while the $|\Psi_2\rangle$ branch has increasing weight.

While we cannot quantitatively evaluate the wave function at long times using the Born-approximated master equation \eqref{eq:master}, the fact that $\lambda<0$ for $w=-1$ means that superpositions of geometries of different cosmological constant will indeed decohere on timescales of order $1/\lambda$, but that the pointer states they decohere into will not be definite-$T_{\mu\nu}$ branches, but instead branches characterized by superpositions of different densities.
Of course, these pointer states could be modified, and the decoherence timescale shortened further, if we introduce the effects of interactions within the matter ($\phi$) sector.

In contrast, for many common values of $w$---e.g., $w=0$ for nonrelativistic matter, $w=1/3$ for incoherent electromagnetic radiation or relativistic matter, etc.---we have $\lambda>0$ and a well-controlled calculation for the decoherence rate~\eqref{eq:lambda} induced by coupling to thermal gravitons.

The generalization from a two-level system to a Hilbert space of arbitrary dimension, and even a continuous distribution of possible $\Phi$ values, is straightforward. 
An off-diagonal element of the density matrix, $\langle \Phi_1 |\rho|\Phi_2\rangle$ for $\Phi_1 \neq \Phi_2$, will evolve according to the master equation \Eq{eq:master3} and in particular will have decoherence rate $\lambda$ going like $f(w)(\Phi_1 - \Phi_2)^2$ as dictated by \Eq{eq:lambda}.
Thus, for positive $f(w)$, the further off the diagonal a density matrix element is, the faster it decoheres.

\subsection{Signs, tensors, decoherence, and causality}\label{sec:causality}

The specific tensor structure appearing in the master equation \eqref{eq:master},
\be
2 T_{\mu\nu}T^{\mu\nu} - T_\mu^{\;\;\mu} T_\nu^{\;\;\nu}, \label{eq:tensorstructure}
\ee
is simply a consequence of the mechanics of integrating over $h_{\mu\nu}$ in \Eq{eq:SIF}.
That is, inverting the two-point function in \Eq{eq:SEgf} and performing the integral over $h_{\mu\nu}$ is closely related to how one integrates out a massive state in vacuum to produce an EFT using the K\"all\'en-Lehmann form of the exact propagator, which for a spin-$s$ state is~\cite{Chandrasekaran:2018qmx}
\be 
\langle \chi_{\mu_1 \cdots \mu_s}(k)\chi_{\nu_1 \cdots \nu_s}(k')\rangle = i(-1)^s \delta^4(k+k') \int_0^\infty {\rm d}\mu^2 \frac{\rho(\mu^2)}{-k^2 - \mu^2 + i\epsilon}\Pi_{\mu_1 \cdots \mu_s\nu_1 \cdots\nu_s}(k),\label{eq:KL}
\ee
where $\Pi$ is the propagator numerator and $\rho$ is the spectral density, which is required by unitarity to be positive.
As in \Ref{Chandrasekaran:2018qmx}, the $(-1)^s$ factor is a consequence of our sign convention for the metric.
In integrating out the thermal environment, we are effectively convolving \Eq{eq:KL} for the massless graviton with its thermal density of states.
From the graviton propagator numerator,
\be
\Pi_{\mu\nu\rho\sigma} = \frac{1}{2}(\overline g_{\mu\rho}\overline g_{\nu\sigma} + \overline g_{\mu\sigma}\overline g_{\nu\rho}) - \frac{1}{2} \overline g_{\mu\nu}\overline g_{\rho\sigma},\label{eq:gravprop}
\ee
one finds that $2T^{\mu\nu}\Pi_{\mu\nu\rho\sigma}T^{\rho\sigma}$ gives precisely the structure in \Eq{eq:tensorstructure}.

Rather than plugging in $T_{\mu\nu}$ for a perfect fluid into the master equation, if we instead plug the Maxwell energy-momentum tensor into \Eq{eq:tensorstructure}, we find that the decoherence rate will go like $2F^{\mu\nu}F_{\nu\rho}F^{\rho\sigma}F_{\sigma\mu} - \frac{1}{2}(F_{\mu\nu}F^{\mu\nu})^2 = \frac{1}{2}[(F_{\mu\nu}F^{\mu\nu})^2 + (F_{\mu\nu}\widetilde F^{\mu\nu})^2]$, which is manifestly nonnegative and, in particular, vanishes for a plane wave of coherent radiation, $F_{\mu\nu}\sim \epsilon_{[\mu}p_{\nu]}$, where $p^2 = \epsilon\cdot p = 0$.
Thus, a coherently polarized beam of light, albeit containing some spectral distribution of wavelengths, does not gravitationally decohere; thermal gravitons do not act like a prism, decohering a beam into its constituent colors on each branch of the wave function. A similar conclusion holds for a plane wave of a massless scalar, but not for a massive scalar.

As we saw in \Sec{sec:OD}, the sign of \Eq{eq:tensorstructure} (equivalently, the sign of $f(w)$) has profound consequences for the evolution of the density matrix, dictating the types of pointer states into which the system decoheres.
In particular, \Eq{eq:tensorstructure}, the form of which derives from the structure of the Lorentzian graviton propagator~\eqref{eq:gravprop}, can be either positive or negative for reasonable $T_{\mu\nu}$.
When $f(w)>0$, the system decoheres into states of definite $\phi$, while if $f(w)<0$, the system evolves into quite different states.
This latter remarkable behavior is uniquely a consequence of the fact that we are coupling the system to a thermal bath of a {\it tensor} field, namely the graviton, which can give rise to a structure like \Eq{eq:tensorstructure}.

Indeed, suppose we replaced the thermal graviton with a massless scalar $\sigma$, so that the interaction term $S_I$ in \Eq{eq:SI} is replaced with $\int {\rm d}^4 x\sqrt{-\overline g} y\sigma\phi$ for some coupling $y$. Since the scalar propagator numerator replacing \Eq{eq:gravprop} is just $1$, we would find that the structure \Eq{eq:tensorstructure} in the master equation would be replaced simply by $2\phi^2$, which is manifestly positive. Then we would simply have the master equation in \Eq{eq:master2}, with $f(w)$ replaced by $2$ and with the $\kappa^2$ in $N$ and $D$ replaced by $y^2$.
We would find that system evolves into the definite-$\phi$ basis.

Similarly, if we have some electromagnetic current coupled to a thermal photon bath, we should replace $S_I$ with $\int {\rm d}^4 x\sqrt{-\overline g} A_\mu J^\mu$.
The propagator numerator for a massless photon is just $\Pi_{\mu\nu} = \overline g_{\mu\nu}$, so for the photon we should replace \Eq{eq:gravprop} with $-\overline g^{\mu\nu}$ and \Eq{eq:tensorstructure} with $-2 J_\mu J^\mu$.
Writing $J_\mu = \Phi u_\mu$ for rest charge density $\Phi$ and fixed comoving four-velocity $u_\mu$, the master equation would appear just as in \Eq{eq:master2} but with $f(w)$ replaced by $-2u_\mu u^\mu$ and with the $\kappa^2$ in $N$ and $D$ replaced by $1$.
That is, the system decoheres into the definite-$\Phi$ basis provided that $J_\mu$ is timelike, i.e., $u_\mu u^\mu = -1$.
Hence, causality of the flow of charge enforces the expected pointer basis for coupling to a photon bath.
In particular, for a system of constant charge density, with possible total charges $Q_1$ and $Q_2$, the decoherence rate is, in analogy with \Eq{eq:lambda}, 
\be 
\lambda = 4\pi T (Q_1-Q_2)^2.
\ee

\section{Discussion}\label{sec:discussion}

We now discuss potential consequences of thermal graviton decoherence in various settings.

\subsection{Consequences for Coleman--De~Luccia}

Let us pause to summarize one of our main findings.
Over the course of the previous three sections, we computed the rate at which a ball of perfect fluid of fixed size $L$, prepared in a superposition of two different energy densities, decoheres under the influence of a thermal bath of gravitons.
For a bath of temperature $T$ and a total energy difference $\Delta E = E_2 - E_1$ between the two states, this decoherence rate is given by \Eq{eq:lambda}.

Let us now suppose that the fluid describes a scalar field with a potential $V(\phi)$ as in \Sec{sec:PT}.
Suppose that the field state is peaked about the false vacuum $\phi_+$, with background temperature as given in \Eq{eq:dStemp} and with a typical density matrix of characteristic energy variance $\Delta E^2 \sim V''(\phi_+)$.
Then, based on the scaling for our ball of size $L$ from \Eq{eq:lambda}, the characteristic timescale over which gravitational interactions drive decoherence is
\be 
t_{\rm dec} \sim \frac{m^3_{\rm Pl}}{V''(\phi_+)\sqrt{V_0}} \sim \left(\frac{m_{\rm Pl}}{m_\varphi}\right)^2 \times\frac{1}{\sqrt{\Lambda}},
\ee
where $V(\phi) \sim \kappa^{-2} \Lambda + \frac{1}{2} m_\varphi^2 (\phi - \phi_+)^2 + \cdots$.
We see that $t_{\rm dec}$ is parameterically longer than the Hubble time $1/\sqrt{\Lambda}$ for reasonable potentials.
However, $t_{\rm dec}$ is nevertheless very short compared to the timescales associated with false vacuum decay (for either Coleman--De~Luccia or Hawking--Moss), which as we saw in \Sec{sec:PT} are exponentially long compared to the characteristic energy scale of the potential.

We conclude that gravitationally-driven decoherence cannot be neglected when calculating the final tunneling rate from false to true vacuum.
However, our calculation does not indicate whether the effect of gravitationally-driven decoherence is to suppress or enhance the tunneling rate.
Moreover, the literature offers circumstantial evidence that can be used to heuristically argue for either possibility, which we briefly outline below.

For the case of suppression, thermal graviton decoherence could stabilize the false vacuum in a quantum Zeno-like way~\cite{Misra:1976by}.
Such behavior occurs in quantum mechanical condensed matter systems, in which dissipative interactions decrease tunneling rates \cite{Caldeira:1982uj}.
For a wave function that is initially peaked about the false vacuum, the cosmological version of the argument is that environmental monitoring by thermal gravitons would reset the spreading wave function on timescales faster than that of Coleman--De~Luccia or Hawking--Moss transitions.
Moreover, excursions away from the false vacuum would be preferentially driven back toward the false vacuum's potential minimum.

For the case of enhancement, graviton decoherence could be the physical mechanism through which the Coleman--De~Luccia bubble nucleation is realized when the wave function has sufficient support over the true vacuum.
In other words, thermal gravitons would ``measure'' the bubble.
A scalar field-based model constructed by Kiefer, Queisser, and Starobinsky supports this possibility, in which they find that decoherence enhances the final vacuum decay rate \cite{Kiefer:2010pb}.

The only way to definitively determine whether gravitationally-driven decoherence enhances or suppresses the rate of vacuum decay would be to compute the final decay rate itself.
In other words, the question is inextricably model-dependent, as one can make heuristic arguments for either possibility.
While our fluid ball model can morally be thought of as a model for a region of space in a superposition of two states---containing only false vacuum and containing a bubble of true vacuum---it only lets us determine the rate at which decoherence alone proceeds.
Calculating the decay rate would require a more elaborate model in which instanton effects were included.
The decay rate will also depend on detailed features of the potential, which affect the state space and pointer states of the gravitational interactions, as well as the choice of initial state.
Nevertheless, attacking the question directly would be an interesting avenue for future research.

\subsection{Experimental consequences}

The decoherence rate derived in \Eq{eq:lambda} relies on the fact that the {\it environment} itself (i.e., the thermal gravitons), in addition to the system that is decohering, is quantum mechanical.
That is, there is nonzero asymptotic decoherence generated by a thermal bath in a Bose--Einstein distribution, but, as we saw in \Sec{sec:classical}, none for a bath following a Maxwell--Boltzmann distribution.
This observation opens the tantalizing possibility of indirectly verifying that gravity is indeed quantized, by observing gravitationally-induced decoherence in the lab.
Given the (in)famous difficulty of directly observing a graviton or verifying that it is quantized~\cite{Eppley1977,Mattingly:2006pu,Dyson,Rothman:2006fp,Ade:2014xna}, a new avenue for such an experiment is valuable, though as we will see, still challenging in practice.

Suppose we take an object of mass $m$ composed of $N$ atoms of atomic mass number $A$ and arrange it in a quantum state with a superposition of nonoverlapping positions. We take as our system states the presence or absence of the mass at a given position. Then the decoherence timescale, setting $w=0$, will be
\be
\lambda^{-1} \simeq  \left(\frac{1\,{\rm K}}{T}\right) N^{-2} A^{-2}\times 1.7\times 10^{20}\,{\rm years} =  \left(\frac{1\,{\rm K}}{T}\right) \left(\frac{1\, {\rm ng}}{m}\right)^2\times 14\,{\rm ms}.
\ee
We expect that there is a cosmological gravitational wave background of temperature of order one Kelvin~\cite{Weinberg:1972kfs}.\footnote{We moreover expect that thermal molecular motion within the earth could generate a graviton environment with temperature of order $10^2$ to $10^3$ K. Analogously, there is a flux of gravitons generated within the sun~\cite{Weinberg:1965nx,Gould}, though whether this competes with the decoherence timescale for cosmic gravitons would involve a more detailed calculation for this type of environment.} Thus, to obtain reasonable laboratory decoherence times as a consequence of cosmological gravitons in the present epoch, one would need to arrange for a superposition of an object containing on the order of $N\sim 10^{10}$ atoms (e.g., for the case of silicon,  where $A=28$).
Arranging for such a mesoscale object to attain a quantum superposition is not outside the realm of possibility for future experiments~\cite{Penrose,Connell2010,RevModPhys.90.025004}.

Another physical consequence one might a priori consider is when the energy difference in $\lambda$ is due to the decay width of some particle state. 
That is, for a particle with lifetime $\tau$, its energy will not be a definite mass eigenstate, but will have width of order $1/\tau$.
If this state is coupled to thermal gravitons, then the decoherence rate $\lambda$ is related to the lifetime by
\be
\lambda \tau \sim \left(\frac{T}{m_{\rm Pl}}\right)\times\left(\frac{t_{\rm Pl}}{\tau}\right) \ll 1,
\ee
since $\tau$ is longer than the Planck time $t_{\rm Pl}$ and we take $T$ less than $m_{\rm Pl}$.
Hence, the particle always decays long before thermal gravitons can project it onto a definite mass eigenstate.

\subsection{Consequences for black holes}

Instead of an unstable particle, let us consider the decay width of a black hole. 
For a black hole of mass $M$, the Hawking temperature $T_{\rm H}=m_{\rm Pl}^2/M$ gives the effective width of the black hole mass.
That is, we can think of the black hole as being in a superposition of different energy eigenstates, with characteristic separation of order $\Delta E \sim T_{\rm H}$.
The Hawking radiation itself generates a thermal bath of gravitons, which probe the black hole geometry and induce back-reaction.
Hence the graviton component of the Hawking radiation acts as a thermal environment, decohering the superposition of black hole mass states in a manner qualitatively similar to that computed quantitatively for our test system.
The timescale of this Hawking radiation-induced decoherence is
\be
\lambda^{-1} \sim T^{-1}_{\rm H}\left(\frac{m_{\rm Pl}}{\Delta E}\right)^2 \sim \frac{M^3}{m_{\rm Pl}^2}.
\ee
Thus, the decoherence timescale is the same as the Page time, in agreement with the branch-counting argument given in \Ref{Bao:2017who}.

\section{Conclusions}\label{sec:conclusions}

In this work we have demonstrated that gravitationally-driven decoherence is both unavoidable and important for the understanding of phase transitions in a cosmological context, particularly in epochs with large de~Sitter temperature. 
In certain situations, the results described in this paper can provide a powerful physical effect that drives cosmological systems away from purity.
In particular, they imply a lower bound on the amount of decoherence happening in our universe at each moment in its history.
It would be a interesting future research direction to better study and categorize such situations, in the contexts of black hole physics, macroscopic superpositions of massive objects, or even conceivably realizable laboratory experiments. 
We leave this rich assortment of potential applications of gravitational decoherence to future research.

\vspace{5mm}

\begin{center}
 {\bf Acknowledgments}
\end{center}
We thank Cliff Burgess, Sean Carroll, Cliff Cheung, Thomas Hertog, Can Kilic, Alex Maloney, Don Marolf, Sonia Paban, and Philippe Sabella-Garnier for useful discussions and comments.  
N.B. is supported by the National Science Foundation under grant number 82248-13067-44-PHPXH,
by the Department of Energy under grant number DE-SC0019380, and by New York State
Urban Development Corporation Empire State Development contract no. AA289.
A.C.-D. is a Postdoctoral Fellow (Fundamental Research) of the Research Foundation - Flanders, File Number 12ZL920N, and he was supported for part of this work by the KU Leuven C1 grant ZKD1118 C16/16/005, the National Science Foundation of Belgium (FWO) grant G.001.12 Odysseus, and by the European Research Council grant no. ERC-2013-CoG 616732 HoloQosmos.
J.P. is supported in part by the Simons Foundation and in part by the Natural Sciences and Engineering Research Council of Canada.
G.N.R. is supported by the Miller Institute for Basic Research in Science at the University of California, Berkeley.

\pagebreak

\appendix

\section{Thermal integrals}\label{app:integrals}

We wish to compute the thermal noise and dissipation integrals ${\cal N}(t)$ and ${\cal D}(t)$ defined in \Eqs{eq:NDdef}{eq:ND}.
Let us commute the ${\bf k}$ and ${\bf r}$ integrals, so that we first evaluate the integral in ${\bf r}$. Since we take $T_{\mu\nu}$ to be a uniform perfect fluid, the only dependence on ${\bf r}$ in the $N$ and $D$ integrands appears through the $e^{i {\bf k}\cdot {\bf r}}$ factor, so the integral over ${\bf r}$ serves to generate the Fourier transform.
We regulate the region of integration to be within the ball ${\cal B}$:
\be
\int_{\cal B} {\rm d}{\bf r}\,e^{i{\bf k}\cdot {\bf r}}  = \frac{4\pi}{k^3}(\sin kL - kL \cos kL) = \frac{4\pi L^2}{k}j_1(kL),
\ee
where, as before, $j_1$ is the first spherical Bessel function of the first kind. We next note that the ${\bf k}$ integrals depend only on the magnitude $k$, so \Eq{eq:ND} becomes:
\be
\begin{aligned}
{\cal N}(t) &=\frac{\kappa^2}{128\pi^3}\int_0^t {\rm d}t'[\mathfrak{N}_0(t') + \mathfrak{N}_1(t')]\\
{\cal D}(t) &= \frac{\kappa^2}{128\pi^3} \int_0^t {\rm d}t' \,\mathfrak{D}(t'),
\end{aligned}\label{eq:calND}
\ee
where
\be
\begin{aligned}
\mathfrak{N}_0(t) &= 16\pi^2 L^2 \int_0^\infty {\rm d}k\,\cos(kt) j_1(kL)\\
\mathfrak{N}_1(t) &= 32\pi^2 L^2 \int_0^\infty {\rm d}k\,\cos(kt)n(k)j_1(kL) \\
\mathfrak{D}(t) &= 16\pi^2 L^2\int_0^\infty {\rm d}k\,\sin(kt)j_1(kL).\label{eq:frakdef}
\end{aligned}
\ee
Performing the $\mathfrak{N}_0$ and $\mathfrak{D}$ integrals, we have
\be 
\mathfrak{N}_0(t) = 16\pi^2 \left\{L+\frac{t}{4}\log\left[\left(\frac{t-L}{t+L} \right)^2 \right]\right\}\label{eq:frakN0}
\ee
and 
\be
\mathfrak{D}(t) = 8\pi^3 t\,\theta(L-t), 
\ee
where $\theta$ is the Heaviside step function.

\begin{figure}[t]
\begin{center}
\includegraphics[width=0.7\columnwidth]{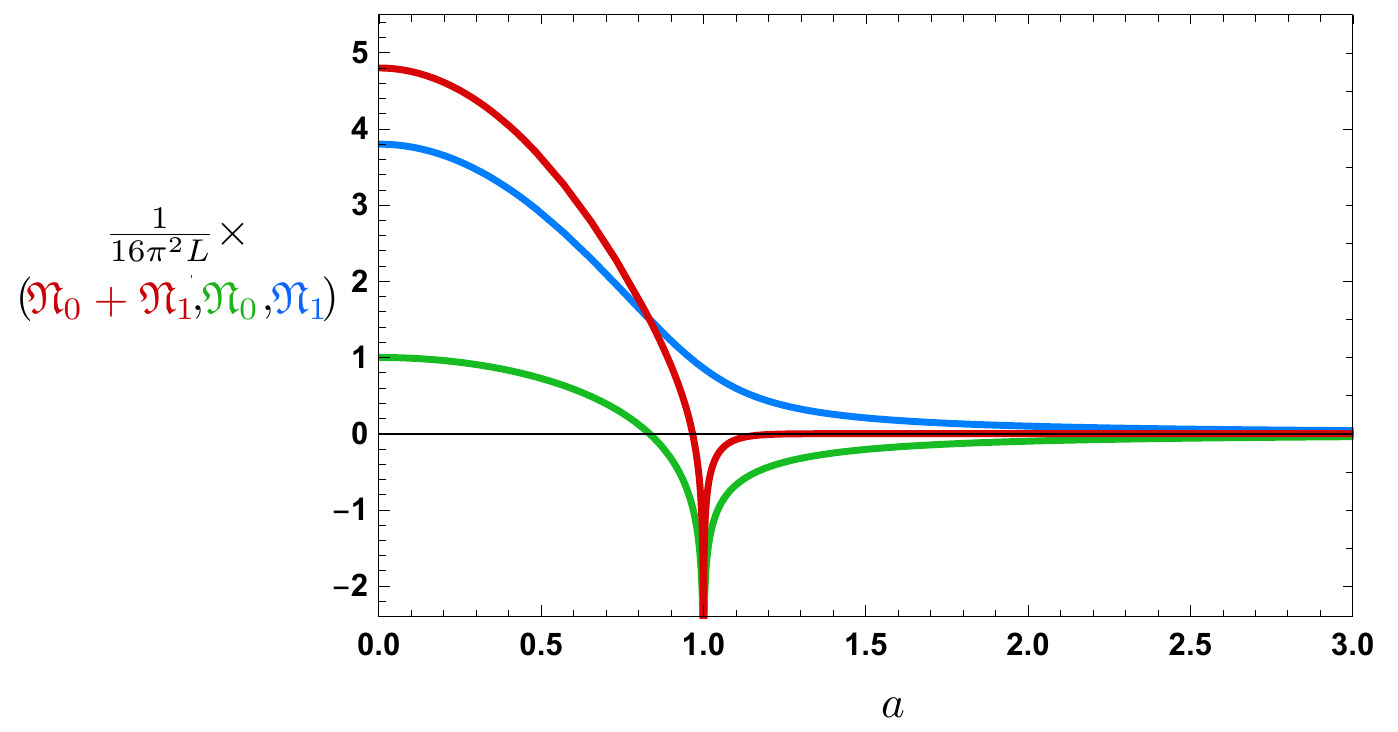}
\end{center}
\vspace{-6mm}
\caption{The noise integrand $\mathfrak{N}_0 + \mathfrak{N}_1$ (red) defined in \Eq{eq:frakdef} as the sum of zero-point part $\mathfrak{N}_0$ (green) from \Eq{eq:frakN0} and Bose--Einstein occupation number contribution $\mathfrak{N}_1$ (blue) from \Eq{eq:frakN1}, as a function of the rescaled time $a=t/L$. In this example, the functions are plotted for rescaled inverse temperature $b = 1/3$.}
\label{fig:appplot}
\end{figure}

We are left with the $\mathfrak{N}_1(t)$ integral, which for $n(k)$ given by the Bose--Einstein distribution in \Eq{eq:BoseEinstein} is highly nontrivial.
Defining unitless variables $x=kL$, $a=t/L$, and $b=\beta/L$, we have 
\be
\begin{aligned}
\frac{\mathfrak{N}_1(t)}{16\pi^2 L} &= 2\int_0^\infty{\rm d}x\,\frac{\cos ax}{e^{bx} - 1}\left(\frac{\sin x}{x^2} - \frac{\cos x}{x} \right)\\
&=\int_0^\infty \frac{{\rm d}x}{e^{bx} - 1} \sum_{n=1}^\infty \frac{(-1)^n x^{2n-1}}{(2n+1)!}[(a-2n)(1+a)^{2n} - (a+2n)(1-a)^{2n}],
\end{aligned}\label{eq:N1def}
\ee
where in the second line we first used trigonometric addition formulas write the numerator in terms of the cosine and sine of $(1\pm a)x$ and then expanded in the Taylor series. Now, the Bose--Einstein integrals are known in closed form,
\be 
\int_0^\infty {\rm d}x\, \frac{x^{2n-1}}{e^{bx}-1} = \frac{\zeta(2n)\Gamma(2n)}{b^{2n}},
\ee
where $\zeta$ is the Riemann zeta function. Using the identity
\be
\zeta(2n) = \frac{(-1)^{n+1} (2\pi)^{2n} B_{2n}}{2(2n!)}, 
\ee
where $B_n$ is the $n$th Bernoulli number, we therefore have
\be 
\frac{\mathfrak{N}_{1}(t)}{16\pi^2 L}=\left[g\left(2\pi\frac{1-a}{b}\right)+g\left(2\pi\frac{1+a}{b}\right)-(1-a)f\left(2\pi\frac{1-a}{b}\right)-(1+a)f\left(2\pi\frac{1+a}{b}\right)\right],\label{eq:N1fg}
\ee 
where we have defined
\be 
\begin{aligned}
f(z) & \equiv\sum_{n=1}^{\infty}\frac{B_{2n}}{4n(2n+1)!}z^{2n}\\
g(z) & \equiv\sum_{n=1}^{\infty}\frac{B_{2n}}{4n(2n)!}z^{2n}.
\end{aligned}\label{eq:fg}
\ee
Rewriting these sums in closed analytic form, we have, within the domain of convergence,\footnote{Using the asymptotic expansion of the Bernoulli numbers, one finds that the radius of convergence of the sums in \Eq{eq:fg} is $|z|<2\pi$, which corresponds to $(1\pm a)<b$ in \Eq{eq:N1fg}. Nonetheless, the full sum of the $f$ and $g$ functions in \Eq{eq:N1fg} converges for {\it all} $a$ and $b$, and the final expression we will obtain for $\mathfrak{N}_1(t)$ matches what one obtains by numerically integrating the defining expression for $\mathfrak{N}_1(t)$ in the first line of \Eq{eq:N1def} for all $a,b$.}
\be 
\begin{aligned}
f(z) & =\frac{1}{4}\left[2-\frac{\pi^{2}}{3z}+\frac{z}{2}-\log z^{2}+\frac{2}{z}{\rm Li}_{2}\left(e^{-z}\right)\right]\\
g(z) & =\frac{1}{2}\log\left[\frac{2}{z}\sinh\left(\frac{z}{2}\right)\right],
\end{aligned}
\ee
where we have used the polylogarithm function ${\rm Li}_n(z) = \sum_{n=1}^\infty z^k/k^n$. Since ${\rm Li}_2(z)\in \mathbb{R}$ for $z\leq 1$ and
\be 
{\rm Li}_{2}(z)+{\rm Li}_{2}(z^{-1})=\frac{\pi^{2}}{3}+i\pi\frac{z}{1-z}\sqrt{\frac{(1-z)^{2}}{z^{2}}}\log z-\frac{1}{2}(\log z)^{2},
\ee
we can define the real part of ${\rm Li}_2(z)$ for all real $z> 0$ as
\be 
\mathfrak{R}{\rm Li}_2(z) = \begin{cases} {\rm Li}_2(z) &z\leq 1 \\ \frac{\pi^2}{3} - \frac{1}{2}(\log z)^2 - {\rm Li}_2(z^{-1}) & z>1  \end{cases},
\ee
which is continuous for all positive $z$.
This allows us to write the manifestly real expression for $\mathfrak{N}_1(t)$ (for $a,b\geq 0$):
\be 
\begin{aligned}
\frac{\mathfrak{N}_{1}(t)}{16\pi^2 L} & =-1+\log2-\frac{\pi(1+a^{2})}{2b}+\frac{\pi b}{12}+\frac{1}{4}a\log\left[\left(\frac{1+a}{1-a}\right)^{2}\right]\\
 & \qquad+\frac{1}{4}\log\left[\sinh^{2}\left(\pi\frac{1-a}{b}\right)\right]+\frac{1}{4}\log\left[\sinh^{2}\left(\pi\frac{1+a}{b}\right)\right]\\
 & \qquad-\frac{b}{4\pi}\left[{\rm Li}_{2}\left(e^{-\frac{2\pi(1+a)}{b}}\right)+\mathfrak{R}{\rm Li}_2\left(e^{-\frac{2\pi(1-a)}{b}}\right)\right].
\end{aligned}\label{eq:frakN1}
\ee
See \Fig{fig:appplot} for an illustration.

Integrating over $t$ (or, equivalently, $a$) as in \Eq{eq:calND}, we have
\be
{\cal D}(t) = 32\kappa^2 L^2 \left[1+(a^{2}-1)\theta(1-a)\right]
\ee
and
\be
{\cal N}(t) = {\cal N}_0(t) + {\cal N}_1 (t), 
\ee
where 
\be
{\cal N}_0(t)= \frac{\kappa^2}{128\pi^3}\int_0^t {\rm d}t'\,\mathfrak{N}_0(t') = \frac{\kappa^2 L^2}{8\pi} \left\{\frac{a}{2} + \frac{1}{8}(1-a)(1+a) \log\left[\left(\frac{1+a}{1-a} \right)^2 \right] \right\}\label{eq:calN0}
\ee
and 
\be
\begin{aligned}
{\cal N}_1 (t) &= \frac{\kappa^2}{128\pi^3}\int_0^t {\rm d}t'\,\mathfrak{N}_1(t')\\&= \frac{\kappa^2 L^2}{8\pi}\biggl\{ \frac{\pi}{3b}-\frac{a}{2}-\frac{1}{8}(1-a)(1+a)\log\left[\left(\frac{1+a}{1-a}\right)^{2}\right]\\
 & \qquad\qquad\qquad+\frac{b}{4\pi}\left[{\rm Li}_{2}\left(e^{-\frac{2\pi(1+a)}{b}}\right)+\mathfrak{R}{\rm Li}_{2}\left(e^{\frac{2\pi(1-a)}{b}}\right)\right]\\
 & \qquad\qquad\qquad+\frac{b^{2}}{8\pi^{2}}\left[{\rm Li}_{3}\left(e^{-\frac{2\pi(1+a)}{b}}\right)-\mathfrak{R}{\rm Li}_{3}\left(e^{\frac{2\pi(1-a)}{b}}\right)\right]\biggr\},
\end{aligned}
\ee
where we use the identity for the trilogarithm (which, like the dilogarithm, is real for real $z\leq 1$),
\be 
{\rm Li}_{3}(z)-{\rm Li}_{3}(z^{-1})=\frac{\pi^{2}}{3}\log z-\frac{1}{6}(\log z)^{3}+\frac{i\pi}{2}\frac{z}{1-z}\sqrt{\frac{(1-z)^{2}}{z^{2}}}(\log z)^{2},
\ee
to define the real part of the trilogarithm for all real $z>0$,
\be 
\mathfrak{R}{\rm Li}_3(z) = \begin{cases} {\rm Li}_3(z) & z\leq 1 \\ {\rm Li}_3(z^{-1}) + \frac{\pi^2}{3} \log z  - \frac{1}{6}(\log z)^3 & z>1 \end{cases},
\ee
which is manifestly real and continuous for all positive $z$.
Thus, we obtain the relatively compact expression for ${\cal N}(t)$:
\be
\begin{aligned}
{\cal N}(t) &= \frac{\kappa^2 L^2}{8\pi}\biggl\{ \frac{\pi}{3b}+\frac{b}{4\pi}\left[{\rm Li}_{2}\left(e^{-\frac{2\pi(1+a)}{b}}\right)+\mathfrak{R}{\rm Li}_{2}\left(e^{\frac{2\pi(1-a)}{b}}\right)\right]\\
 & \qquad\qquad\;\;\;\;\,\,+\frac{b^{2}}{8\pi^{2}}\left[{\rm Li}_{3}\left(e^{-\frac{2\pi(1+a)}{b}}\right)-\mathfrak{R}{\rm Li}_{3}\left(e^{\frac{2\pi(1-a)}{b}}\right)\right]\biggr\}.
\end{aligned}
\ee
See \Fig{fig:N}.

\noindent

\bibliographystyle{utphys-modified}
\bibliography{GravitonDecoherence}

\end{document}